\documentclass[12pt]{article}

\usepackage{geometry}
\usepackage{amsmath}
\usepackage{amssymb}
\usepackage{amsxtra}
\usepackage{amsthm}
\usepackage{mathrsfs}
\usepackage{mathtools}
\usepackage{graphicx}
\usepackage{epstopdf}
\usepackage{subfigure}
\usepackage{float}
\usepackage{xcolor,graphicx,float}
\usepackage[titletoc]{appendix}
\usepackage[colorlinks,linkcolor=red,anchorcolor=blue,citecolor=blue]{hyperref}
\usepackage{enumerate}
\usepackage{verbatim}
\usepackage{bm}
\usepackage{ragged2e} 
\usepackage{booktabs,makecell, multirow, tabularx}
\usepackage{cancel}

\usepackage{natbib}  
 
\def\EquationsBySection{\def\theequation
{\thesection.\arabic{equation}}%
\@addtoreset{equation}{section}}

\usepackage{indentfirst}
\newtheorem{remark}{Remark}[section]

\newtheorem{condition}[remark]{Assumption}
\newtheorem{proposition}[remark]{Proposition}

\newtheorem{corollary}[remark]{Corollary}
\newtheorem{lemma}[remark]{Lemma}

\newcommand\old[1]{}
\makeatother
\EquationsBySection

\renewcommand{\geq}{\geqslant}
\renewcommand{\leq}{\leqslant}

\newcommand{\R}{\mathbb{R}}

\newcommand{\bbm}{\begin{bmatrix}}
\newcommand{\ebm}{\end{bmatrix}}

\renewcommand{\geq}{\geqslant}
\renewcommand{\leq}{\leqslant}
\renewcommand{\ge}{\geqslant}
\renewcommand{\le}{\leqslant}

\newcommand{\dd}{\operatorname{d}\! }

\newcommand{\ds}{\operatorname{d}\! s}
\newcommand{\dt}{\operatorname{d}\! t}
\newcommand{\du}{\operatorname{d}\! u}

\newcommand{\dw}{\operatorname{d}\! W}

\newcommand{\dz}{\operatorname{d}\! z}

\begin{document}
\pagestyle{plain}
\title{Dynamic Portfolio Selection under Range Value-at-Risk
with Logarithmic Returns}
 
\author{
Hui Mi \thanks{School of Mathematical Sciences,
Nanjing Normal University, Nanjing, China. Email: \url{mihui@njnu.edu.cn}. }
\and
Yuzhe Pan\thanks{School of Mathematical Sciences,
Nanjing Normal University, Nanjing, China. Email: \url{230902016@njnu.edu.cn}. }
\and
Zuo Quan Xu\thanks{Department of Applied Mathematics, The Hong Kong Polytechnic University, Kowloon, Hong Kong SAR, China. Email: \url{maxu@polyu.edu.hk}. 
}}
\date{}

\maketitle

\begin{abstract} 
This paper studies continuous-time portfolio selection under Range Value-at-Risk (RVaR), with both portfolio performance and risk evaluated in terms of logarithmic returns. We formulate a mean--RVaR
problem in a continuous-time financial market and reduce the dynamic optimization to a static quantile problem. Using quantile representation, Lagrangian optimization, and convexification, we
characterize the optimal terminal wealth through the convex envelope of an RVaR-specific transformed weighting function.

The RVaR-specific convex envelope exhibits two distinct geometric configurations. Under Scenario~1, the terminal wealth has a VaR-type structure: the minimum-RVaR payoff is digital-option-like, whereas for
$0<\lambda<1$ it consists of two growth-optimal regions separated by a constant-payoff region. Under Scenario~2, the convex envelope is determined either by a supporting line connecting an interior contact
point to the endpoint or by a common supporting line with two interior contact points, leading to a richer multi-region payoff structure. We further show that the RVaR solutions reduce to their VaR
and Expected Shortfall (ES) counterparts as $\beta\downarrow\alpha$ and $\alpha=0$, respectively. Under constant investment opportunities, we also derive closed-form wealth processes and replicating portfolio
strategies.

Numerical results show that the effects of $(\alpha,\beta,\lambda)$ on risky-asset exposure are nonlinear
and generally non-monotone. In particular, when $\alpha>0$, pure minimum-RVaR optimization may
assign zero terminal wealth in sufficiently bad states because the extreme quantile region $[0,\alpha)$ is excluded from the RVaR criterion.
These results clarify how interval-based quantile risk control shapes state-dependent portfolio payoffs and connects RVaR with its VaR and ES limiting cases.

\end{abstract}

\noindent
\textbf{Keywords:}
Range Value-at-Risk; logarithmic return; quantile optimization; convex envelope; portfolio selection.

\section{Introduction}
Uncertainty and risk are inherent in financial markets, and the
measurement and management of financial risk have long been central
topics in portfolio theory. Since   \cite{Markowitz}
introduced modern portfolio theory, a fundamental objective of
portfolio selection has been to balance expected return against risk
through asset allocation and diversification. Among the earliest and
most widely used risk measures, variance quantifies the dispersion of
returns around their mean \cite{He1,Jin2,Jin3}. Its symmetric
treatment of positive and negative deviations, however, is not fully
aligned with investors' primary concern with downside risk.

To address this limitation, downside-oriented measures such as
semi-variance and semi-absolute deviation have been proposed
\cite{Jin1}. These measures focus on unfavorable deviations, but they
do not directly identify economically meaningful loss thresholds.
Value-at-Risk (VaR) subsequently became one of the most widely used
quantile-based risk measures because of its intuitive interpretation
and comparability across portfolios \cite{Wei1}. For a given
probability level, VaR identifies a lower-tail loss threshold rather
than the maximum possible loss. As a single-quantile criterion,
however, VaR does not reflect the severity of losses beyond the
specified quantile. Moreover, VaR is not subadditive in general, and
therefore does not always reward diversification in the manner
required by coherent risk measurement \cite{Danielsson}.

Expected Shortfall (ES), also referred to as Conditional
Value-at-Risk (CVaR) in part of the literature, provides a natural
alternative by averaging losses over the tail beyond a prescribed
quantile level \cite{Rockafellar1,Rockafellar2,Wei2}. In its quantile
representation, ES incorporates information from the entire lower tail
and is a coherent risk measure. This makes ES particularly suitable
for applications in portfolio optimization, capital allocation, and
tail-risk management. More general frameworks, including spectral risk
measures \cite{Acerbi} and distortion risk measures, have further
extended the class of law-invariant tail-risk measures. Related
continuous-time portfolio problems have also been studied extensively
in the literature \cite{Basak}.

In addition to coherence, the statistical robustness of risk
measurement procedures has received considerable attention
\cite{Cont,Kou}. \cite{Cont} emphasized
that a risk measure may be theoretically appealing yet sensitive to
small perturbations in the underlying distribution. The tension
between coherence and statistical robustness has therefore become an
important consideration in the design and selection of risk measures.
Kou and Peng \cite{Kou} compared the robustness properties of VaR and ES,
while  \cite{Danielsson} discussed conditions under
which VaR may retain desirable diversification properties in
applications.  \cite{Ibragimov} further showed that
under sufficiently heavy-tailed risks, diversification need not reduce
tail risk. These results suggest that robustness and tail sensitivity,
rather than coherence alone, are both relevant for practical risk
management.

Range Value-at-Risk (RVaR), introduced by   \cite{Cont},
provides a flexible intermediate risk measure between a single quantile and
an average over a range of lower-tail quantiles. For $0\leq\alpha<\beta\leq1$, RVaR averages the
quantile function over the interval $[\alpha,\beta]$. 
Hence, it focuses on a prescribed range of lower-tail quantiles rather
than on a single quantile or an average over the entire lower tail. 
In particular, RVaR
reduces to ES when $\alpha=0$, while it converges to VaR as
$\beta\downarrow\alpha$ under the quantile convention adopted in this
paper. This two-parameter structure allows the risk manager to control
both the location and the width of the quantile region entering the
risk criterion. RVaR has been studied in forecasting, risk allocation, and reinsurance;
see, for example, \cite{Maria,Wang,Zhang}. Its robustness properties provide an additional motivation for its use in situations
where model misspecification is a concern.

Model uncertainty provides another motivation for studying RVaR.
In many financial applications, the exact distribution of the
underlying risks is unknown and only partial distributional
information, such as moments or shape restrictions, may be available.
Under such circumstances, worst-case analysis evaluates the supremum
of a risk measure over all distributions consistent with the available
information and therefore provides a natural tool for stress testing
and conservative risk assessment.  \cite{Li} derived
explicit worst-case RVaR bounds under moment and shape constraints for
both individual and aggregate risks, thereby providing a theoretical
benchmark for RVaR under distributional ambiguity.

Despite these developments, incorporating law-invariant quantile risk
measures into continuous-time portfolio optimization remains
technically challenging.  \cite{He2} showed that, in a
complete continuous-time market, mean-risk models based on Weighted
Value-at-Risk (WVaR) of terminal wealth may become ill-posed and lead
to extreme leverage. In such cases, the usual economic trade-off
between expected terminal wealth and risk may degenerate.
 
 \cite{Wei3} addressed this difficulty by shifting the
object of both performance evaluation and risk measurement from
terminal wealth to logarithmic return. This choice is closely related
to the Growth Optimal Portfolio (GOP) and the Kelly criterion
\cite{Kelly}, since expected logarithmic wealth growth is the natural
objective in growth-optimal portfolio theory. More importantly, the
logarithmic transformation introduces concavity into the objective and
makes quantile optimization and convexification techniques applicable.
The resulting optimal terminal wealth typically exhibits
state-dependent regions in which wealth is either proportional to the
GOP payoff or remains constant.

Despite the general WVaR formulation in  \cite{Wei3}, the
RVaR case has not been characterized explicitly. Unlike VaR or ES,
RVaR assigns weight to an interior quantile interval
$[\alpha,\beta]$, which generates a transformed weighting function
with multiple slope regimes. Consequently, its convex envelope may
exhibit qualitatively different contact structures, and the resulting
optimal terminal wealth cannot be obtained by a direct substitution
into the existing VaR or ES solutions.
Motivated by these results, this paper studies a continuous-time
portfolio selection problem in which both expected performance and
RVaR are evaluated in terms of logarithmic return. Specifically, for
\[
R_T=\frac{1}{T}\log\frac{X_T}{x},
\]
we consider the mean--RVaR criterion
\[
\lambda\mathbb E[R_T]
-
(1-\lambda)\operatorname{RVaR}_{\alpha,\beta}(R_T),
\qquad
0\leq\lambda\leq1.
\]
The parameter $\lambda$ determines the relative weight assigned to
expected logarithmic growth and RVaR. The case $\lambda=0$
corresponds to pure minimum-RVaR optimization, whereas $\lambda=1$
reduces to the growth-optimal benchmark.

The main difficulty is that, after reducing the dynamic portfolio
problem to a static terminal-wealth problem, the decision variable is
a quantile function and must therefore satisfy a monotonicity
constraint. Consequently, the corresponding Lagrangian cannot in
general be maximized pointwise. We overcome this difficulty by
combining a state-price-density rearrangement, a quantile
transformation, and convex-envelope techniques. For RVaR, the
resulting transformed weighting function has a characteristic
three-region structure determined by the quantile interval
$[\alpha,\beta]$. Its convex envelope generates two distinct geometric
scenarios and, correspondingly, different optimal terminal-wealth
structures.

The main contributions of this paper are as follows.
 
First, building on the log-return WVaR framework of \cite{Wei3},
we derive explicit RVaR-specific solutions for the optimal terminal
wealth. The interior quantile interval $[\alpha,\beta]$ generates two
distinct convex-envelope geometries. Under Scenario~1, the
minimum-RVaR solution has a digital-option-type payoff, whereas for
$0<\lambda<1$ the optimal terminal wealth consists of two
GOP-type regions separated by a constant-payoff region. Under
Scenario~2, the convex envelope is determined either by a supporting
line reaching the endpoint or by a common supporting line with two
interior contact points, resulting in a richer multi-region wealth
structure.

Second, using the quantile optimization approach
\cite{Embrechts,He3,Xu2}, we provide
a unified derivation of the optimal terminal wealth and, under
constant investment opportunities, derive the corresponding wealth
processes and replicating portfolio strategies.

Third, we establish the limiting connections between RVaR, VaR, and
ES. As $\beta\downarrow\alpha$, the RVaR criterion and the associated
optimal terminal wealth reduce to their VaR counterparts. When
$\alpha=0$, RVaR coincides with ES and the Scenario~2 solution reduces
to the corresponding ES structure. These limiting cases provide an
internal consistency check for the general RVaR formulation.

Finally, the numerical analysis illustrates how the parameters
$(\alpha,\beta,\lambda)$ affect the optimal terminal wealth and the
initial risky-asset exposure. The comparative statics are nonlinear
and generally non-monotone: neither the interval width
$\beta-\alpha$ nor the weight $\lambda$ induces a globally monotone
change in risky-asset exposure. In addition, when $\alpha>0$, the
extreme quantile region $[0,\alpha)$ is excluded from the RVaR
criterion. Consequently, in the pure minimum-RVaR problem, the
optimizer may assign zero terminal wealth to sufficiently adverse
states, highlighting an important implication of localized
quantile-risk optimization.

The remainder of the paper is organized as follows. Section~\ref{sec:pre}
introduces the relevant risk measures and the continuous-time
financial market. Section~\ref{sec:strategy_RVaR} formulates the mean--RVaR portfolio
problem, develops its quantile representation, derives the optimal
terminal wealth under the two scenarios, and studies the limiting VaR
and ES cases. Section~\ref{sec:numerical} presents numerical illustrations and
comparative statics. Section~\ref{sec:conclude} concludes. Technical proofs are
provided in Appendices \ref{A.1}-\ref{D.1}.

\section{Risk Measures and Financial Market}\label{sec:pre}
\subsection{Notation} \label{sec:pre1}
Let $(\Omega, \mathbb{F},\mathbb P)$ be a probability space, and let
$
L_B:=L_B(\Omega,\mathbb{F},\mathbb P)
$
denote the set of all finite-valued random variables that are bounded
from below almost surely. Throughout the paper, a random variable
$X\in L_B$ is interpreted as the profit and loss (P\&L) of an
investment, with positive values representing gains and negative
values representing losses.

Let $\mathcal F$ denote the set of all cumulative distribution functions (CDFs) associated
with random variables in $L_B$, that is,
\[
\mathcal F
=
\left\{
F:\mathbb R\to[0,1]:
\begin{array}{l}
F \text{ is nondecreasing and right-continuous},\\
F(a)=0 
\text{ for some } a\in\mathbb R,
\displaystyle\lim_{x\to\infty}F(x)=1
\end{array}
\right\}.
\]
For $F\in\mathcal F$, we adopt the right-continuous quantile
convention and define
\[
F^{-1}(u)
:=
\inf\{x\in\mathbb R:F(x)>u\}
=
\sup\{x\in\mathbb R:F(x)\le u\},
\qquad u\in[0,1).
\]
Let
\(
\mathcal G
:=
\left\{
F^{-1}:F\in\mathcal F
\right\}
\)
be the corresponding class of quantile functions. Equivalently,
\[
\mathcal G
=
\left\{
G:[0,1)\to\mathbb R:
G \text{ is nondecreasing and right-continuous}
\right\}.
\]
For convenience, we set $G(1):=\lim_{t\to 1-}G(t)\in \R\cup\{+\infty\}$. For a random variable $X$,
we denote its distribution and quantile functions by $F_X$ and $G_X$,
respectively. We have $X\in L_B$ if and only if $G_X\in \mathcal G$.

\subsection{Risk Measures} \label{RM}
We now introduce the risk measure Weighted Value-at-Risk (WVaR), which provides the
general framework within which RVaR is defined.

\subsubsection{Properties of Risk Measures}

Following the risk-measure framework used in the weighted-VaR
literature, consider a functional
\[
\rho:L_B\to(-\infty,\infty],
\]
i.e.\ $\rho$ is allowed to take the extended value $+\infty$. This occurs,
for example, for the WVaR of a logarithmic return in
Section~\ref{subsec:RVaR-log-return}.
We recall the following properties.

\begin{enumerate}
\item {Monotonicity.}
For any $X,Y\in L_B$ such that $X\le Y$ almost surely,
\[
\rho(X)\ge \rho(Y).
\]

\item {Cash invariance (translation invariance).}
For any $X\in L_B$ and constant $a\in\mathbb R$,
\[
\rho(X+a)=\rho(X)-a.
\]

\item {Truncation continuity.}
For any $X\in L_B$,
\[
\rho(X)=\lim_{n\to\infty}\rho(X\wedge n).
\]

\item {Positive homogeneity.}
For any $X\in L_B$ and constant $c>0$,
\[
\rho(cX)=c\rho(X).
\]

\item {Subadditivity.}
For any $X,Y\in L_B$,
\[
\rho(X+Y)\le \rho(X)+\rho(Y).
\]

\item {Convexity.}
For any $X,Y\in L_B$ and constant $\gamma\in[0,1]$,
\[
\rho\bigl(\gamma X+(1-\gamma)Y\bigr)
\le
\gamma\rho(X)+(1-\gamma)\rho(Y).
\]

\item {Law invariance.}
If $X,Y\in L_B$ have the same distribution, then
\[
\rho(X)=\rho(Y).
\]

\item {Comonotonic additivity.}
If $X,Y\in L_B$ are comonotonic, that is,
\[
\bigl(X(\omega)-X(\omega')\bigr)
\bigl(Y(\omega)-Y(\omega')\bigr)\ge0,
\qquad
\forall\,\omega,\omega'\in\Omega_0,
\]
for some $\Omega_0\subseteq \Omega$ with full measure $\mathbb P(\Omega_0)=1$, then
\[
\rho(X+Y)=\rho(X)+\rho(Y).
\]
\end{enumerate}

In the terminology adopted in this paper, a functional satisfying
Properties~1--3 is referred to as a risk measure. A risk measure that
also satisfies positive homogeneity and subadditivity is referred to
as coherent. See  \cite{Wei1} for further details.
\subsubsection{VaR, ES, WVaR, and RVaR}

For a P\&L random variable $X\in L_B$ and a lower-tail probability
level $\alpha\in(0,1)$, the VaR of $X$ at level $\alpha$ is defined by
\[
\operatorname{VaR}_{\alpha}(X)
:=
-G_X(\alpha).
\]
Thus, under our P\&L convention, VaR is the negative of the
$\alpha$-quantile of $X$. Although VaR provides a simple and
interpretable quantile-based measure of downside risk, it does not
incorporate information on losses beyond the selected quantile and is
not subadditive in general.

Expected Shortfall (ES) incorporates information over the entire lower
tail. At level $\alpha\in(0,1)$, it is defined by
\[
\operatorname{ES}_{\alpha}(X)
:=
-\frac{1}{\alpha}
\int_0^\alpha G_X(u)\du.
\]
ES is coherent \cite{Artzner,Rockafellar1}, law-invariant, and
comonotonic additive \cite{Acerbi}.

A convenient framework encompassing both quantile and tail-average
risk measures is Weighted Value-at-Risk (WVaR). Let
$\mathcal P([0,1])$ denote the set of probability measures on
$[0,1]$. For $\Phi\in\mathcal P([0,1])$, define its associated WVaR as 
\begin{equation}
\rho_{\Phi}(X)
=
-\int_{[0,1]}G_X(u)\,\Phi(\du).
\label{eq:WVaR}
\end{equation}

\begin{itemize}
\item 
If $\Phi$ is the Dirac probability measure concentrated
at $\alpha$, then
\(
\rho_{\Phi}(X)=\operatorname{VaR}_{\alpha}(X).
\)
\item 
If $\Phi$ is absolutely continuous with density
\(
\phi_\alpha(u)
=
\frac{1}{\alpha}\mathbf 1_{[0,\alpha]}(u),
\)
then
\(
\rho_{\Phi}(X)=\operatorname{ES}_{\alpha}(X).
\)
\end{itemize}
WVaR is law-invariant and comonotonic additive and provides a general
quantile-based framework for risk measurement.

Although ES incorporates information from the entire lower tail and is
coherent, its statistical robustness has been questioned in the literature
\cite{Cont,Kou}. Indeed, ES is an example of the more general conflict between coherence and robustness. Intuitively, a robust risk measure is less sensitive to outliers, and slight misspecifications of the econometric model do not lead to significant differences in risk estimates and forecasts.  \cite{Cont} addressed this lack of robustness by proposing RVaR, a risk measure that averages quantiles over a prescribed range.

For $0\le\alpha<\beta\le1$, define
\[
\operatorname{RVaR}_{\alpha,\beta}(X)
=
-\frac{1}{\beta-\alpha}
\int_{\alpha}^{\beta}G_X(u)\du.
\]
The pair $(\alpha,\beta)$ determines the lower-tail region over which
quantiles are averaged. The parameter $\alpha$ specifies the lower
endpoint of the evaluated range, $\beta$ specifies its upper endpoint,
and $\beta-\alpha$ determines the width of the averaging window which can be understood as the granularity of risk control. A narrower interval implies more precise risk control (similar to VaR); a wider interval implies smoother risk control (similar to ES). .

Two important special or limiting cases are
\[
\operatorname{RVaR}_{0,\beta}(X)
=
\operatorname{ES}_{\beta}(X),
\]
and, under the right-continuous quantile convention adopted above,
\[
\lim_{\beta\downarrow\alpha}
\operatorname{RVaR}_{\alpha,\beta}(X)
=
\operatorname{VaR}_{\alpha}(X).
\]
For $\alpha>0$, RVaR is not subadditive in general and is therefore
not coherent. In the present positively homogeneous setting, this
failure of subadditivity also precludes convexity.

\subsection{Financial Market} \label{FM}

Let $T>0$ be a fixed investment horizon. Consider a filtered
probability space
$
(\Omega, \mathbb{F},(\mathcal F_t)_{0\le t\le T},\mathbb P),$
supporting an $n$-dimensional standard Brownian motion
$
W_t=(W_t^1,\ldots,W_t^n)^\top.
$
Assume $\mathcal{F}_t=\sigma\{W_s:0\le s\le t\}$, augmented by all $\mathbb{P}$-null sets in $\mathbb{F}$.

Following   \cite{Karatzas}, we consider a continuous-time financial market with $n+1$ continuously traded assets. The price process of the risk-free asset satisfies
\[
\dd S_t^0=r_tS_t^0\dt,
\qquad
S_0^0=s_0>0,
\]
where $r_t$ is a uniformly bounded, progressively measurable
real-valued process. For $i=1,\ldots,n$, the price of the $i$th risky asset satisfies
\[
\dd S_t^i
=
S_t^i
\left(
\mu_t^i\dt
+
\sum_{j=1}^n\sigma_t^{ij}\dw_t^j
\right),
\qquad
S_0^i=s_i>0,
\]
where $\mu_t^i$ and $\sigma_t^{ij}$ are progressively measurable and
satisfy the required integrability conditions.

Define
\[
\mu_t
=
(\mu_t^1,\ldots,\mu_t^n)^\top,
\qquad
\sigma_t
=
(\sigma_t^{ij})_{n\times n},
\]
and let
\[
B_t
:=
\mu_t-r_t\mathbf 1
=
(\mu_t^1-r_t,\ldots,\mu_t^n-r_t)^\top
\]
denote the vector of instantaneous excess returns.

To ensure that the state-price density is well defined and to support
the subsequent quantile formulation, we impose the following
assumptions on the market parameters.
\begin{condition}
\label{ass:market}
The volatility matrix $\sigma_t$ is nonsingular $\mathbb P$-a.s. for
Lebesgue-a.e. $t\in[0,T]$. Define the market price of risk by
\(
\theta_t:=\sigma_t^{-1}B_t.
\)
Assume that $\theta$ is progressively measurable and satisfies the Novikov condition 
\[
\mathbb E
\left[
\exp\left(
\frac12\int_0^T\|\theta_t\|^2\dt
\right)
\right]
<\infty.
\]
\end{condition}

Assumption~\ref{ass:market} ensures that the market price of risk is
well defined and that the stochastic exponential associated with
$\theta$ is a true martingale. In the present $n$-stock, $n$-Brownian setting, together with the stated
integrability conditions, the nonsingularity of $\sigma_t$ yields market
completeness.

Consider an investor with initial wealth $x>0$. A portfolio strategy
is an $\mathbb R^n$-valued progressively measurable process
\(
\pi_t=(\pi_t^1,\ldots,\pi_t^n)^\top,
\)
where $\pi_t^i$ denotes the amount invested in the $i$th risky asset
at time $t$. Under continuous self-financing trading with no
transaction costs, the corresponding wealth process satisfies
\[
\dd X_t
=
\left(r_tX_t+B_t^\top\pi_t\right)\dt
+
\pi_t^\top\sigma_t\dw_t,
\qquad
X_0=x.
\]
A portfolio strategy $\pi$ is called admissible if it is progressively
measurable, satisfies 
\[
\int_0^T
\|\sigma_t^\top\pi_t\|^2\dt<\infty,
\qquad
\int_0^T
|B_t^\top\pi_t|\dt<\infty,
\quad\text{a.s.},
\]
and its corresponding terminal wealth satisfies $X_T\ge0$, a.s.

Under Assumption~\ref{ass:market}, define the state-price density
process by
\begin{equation} 
\xi_t
=
\exp\left\{
-\int_0^t
\left(
r_s+\frac12\|\theta_s\|^2
\right)\ds
-\int_0^t\theta_s^\top\dw_s
\right\}.
\label{2.2} 
\end{equation}
Then
\(
0<\xi_T<\infty~~
\) a.s. Because $r$ is uniformly bounded and the stochastic exponential
satisfies the above integrability condition,
\(
0<\mathbb E[\xi_T]<\infty.
\)

Market completeness allows the dynamic portfolio problem to be
reformulated as a static terminal-wealth problem. In particular, any
attainable nonnegative terminal wealth must satisfy the budget
constraint
\[
\mathbb E[\xi_TX_T]\le x.
\]
Conversely, subject to the admissibility conditions imposed above,
an appropriate terminal payoff satisfying the budget constraint can
be replicated by a self-financing trading strategy. This static
representation will be used in the quantile formulation developed in
the next section.

\section{Problem Formulation and Optimal Strategy} \label{sec:strategy_RVaR}

\subsection{RVaR of Logarithmic Return and Problem Formulation}
\label{subsec:RVaR-log-return}

We first introduce the performance variable used throughout the
portfolio optimization problem. Let
\[
R_T
:=
\frac{1}{T}\log\frac{X_T}{x},
\]
where $x>0$ is the initial wealth and $X_T\ge0$ is the terminal
wealth. Thus, $R_T$ represents the average continuously compounded
rate of return over the investment horizon $[0,T]$.

Because admissible terminal wealth is allowed to vanish, we adopt the
extended-real convention
\[
\log0:=-\infty,
\qquad
e^{-\infty}:=0.
\]
Accordingly, $R_T$ may take values in $[-\infty,\infty)$.
In particular, if
\(
\mathbb P(X_T=0)>0,
\)
then
\(
\mathbb E[R_T]=-\infty
\)
in the extended-real sense.

Since the logarithm is strictly increasing on $(0,\infty)$, and under
the right-continuous quantile convention introduced in
Section~\ref{RM}, the quantile function of $R_T$ is given by
\[
G_{R_T}(u)
=
\frac{1}{T}
\log\frac{G_{X_T}(u)}{x},
\qquad
u\in[0,1].
\]

Maximizing $\mathbb E[R_T]$ is equivalent to maximizing expected
logarithmic terminal wealth and therefore leads, in the absence of
risk control, to the logarithmic-utility or growth-optimal portfolio.

We first consider the WVaR of $R_T$. For
$\Phi\in\mathcal P([0,1])$, using the definition of WVaR and the fact
that $\Phi$ is a probability measure, we obtain
\begin{align}
\rho_\Phi(R_T)
&=
-\int_{[0,1]}G_{R_T}(u)\,\Phi(\du)
\nonumber\\
&=
-\frac{1}{T}
\int_{[0,1]}
\log\frac{G_{X_T}(u)}{x}\,\Phi(\du)
\nonumber\\
&=
-\frac{1}{T}
\int_{[0,1]}
\log G_{X_T}(u)\,\Phi(\du)
+
\frac{1}{T}\log x.
\label{eq:WVaR-log-return}
\end{align}

The logarithmic integral in
\eqref{eq:WVaR-log-return} requires special treatment when the terminal
wealth can vanish. Define
\[
\mathcal G_\Phi
:=
\left\{
G\in\mathcal G:
G(s)>0
\ \text{whenever}\
\Phi([0,s])>0,
\quad s\in[0,1]
\right\}.
\]
For $G\in\mathcal G\setminus\mathcal G_\Phi$, we adopt the convention
\[
\int_{[0,1]}
\log G(u)\,\Phi(\du)
:=
-\infty.
\]
Consequently, the corresponding WVaR of logarithmic return satisfies
\[
\rho_\Phi(R_T)=+\infty.
\]
Thus, if zero terminal wealth occurs over a range of quantile levels
receiving positive $\Phi$-mass, the resulting logarithmic-return risk
is infinite.

For RVaR with $0\le\alpha<\beta\le1$, the corresponding weighting
measure is the uniform probability measure on $[\alpha,\beta]$, namely,
\[
\Phi_{\alpha,\beta}(\du)
=
\frac{1}{\beta-\alpha}
\mathbf 1_{[\alpha,\beta]}(u)\du.
\]
Equivalently, its distribution function is
\begin{equation}
\Phi_{\alpha,\beta}([0,u])
=
\begin{cases}
0,
& u<\alpha,\\[1mm]
\dfrac{u-\alpha}{\beta-\alpha},
& \alpha\le u\le\beta,\\[2mm]
1,
& u>\beta.
\end{cases}
\label{3.2} 
\end{equation}
Therefore,
\begin{align}
\operatorname{RVaR}_{\alpha,\beta}(R_T)
&=
-\frac{1}{\beta-\alpha}
\int_\alpha^\beta
G_{R_T}(u)\du
\nonumber\\
&=
-\frac{1}{T(\beta-\alpha)}
\int_\alpha^\beta
\log G_{X_T}(u)\du
+
\frac{1}{T}\log x.
\label{eq:RVaR-log-return}
\end{align}

We now formulate the portfolio optimization problem. For
$0<\lambda<1$, consider the mean--RVaR criterion maximizing problem: 
\begin{align}\label{3.3}
\max_\pi\quad &\lambda\mathbb{E}[R_T]-(1-\lambda)\operatorname{RVaR}_{\alpha,\beta}(R_T)\notag\\
\mathrm{s.t.}\quad&\dd X_t=(r_tX_t+\pi_t^\top B_t)\dt+\pi^\top_t\sigma_t\dw_t,~~
X_T\ge 0,~~X_0=x.
\end{align}
Here $\lambda$ is a preference parameter governing the trade-off
between expected logarithmic growth and RVaR risk.

By market completeness and the terminal-wealth representation
established in Section~\ref{FM}, the dynamic problem can be reformulated as
the following static problem:
\begin{align} \label{3.4}
\max_{X_T}\quad &\lambda\mathbb{E}[R_T]-(1-\lambda)\operatorname{RVaR}_{\alpha,\beta}(R_T) \notag\\
\mathrm{s.t.}\quad &\mathbb{E}[\xi_T X_T]\le x,~ X_T\ge 0,
\end{align}
where $\xi_T$ is given by (\ref{2.2}). Once the optimal terminal
wealth is obtained, the corresponding dynamic portfolio strategy can
be recovered by replication. 

The two boundary cases $\lambda=0$ and $\lambda=1$ are treated
separately. 

When $\lambda=1$, the problem becomes
\begin{align}\label{3.6}
\max_{X_T}\quad &\mathbb{E}[R_T]\notag\\
\mathrm{s.t.}\quad&\mathbb{E}[\xi_T X_T]\le x, ~X_T\ge 0,
\end{align}
which is the classical logarithmic-utility, or growth-optimal,
portfolio problem. 
The optimal terminal wealth in (\ref{3.6}) is that generated by the
growth-optimal portfolio (GOP), or Kelly strategy \cite{Kelly}, and is given
by
\begin{align}\label{3.7}
X_T^{\mathrm{GOP}}=\frac{x}{\xi_T}.
\end{align}
The corresponding wealth process is
\[
X_t^{\mathrm{GOP}}
=
\frac{x}{\xi_t},
\]
and the associated optimal portfolio is
\[
\pi_t^{\mathrm{GOP}}
=
X_t^{\mathrm{GOP}}
(\sigma_t^\top)^{-1}\theta_t
=
\frac{x}{\xi_t}
(\sigma_t^\top)^{-1}\theta_t.
\]

When $\lambda=0$, the problem reduces to the pure minimum-RVaR
problem
\begin{align}\label{3.5}
\min_{X_T}\quad &\operatorname{RVaR}_{\alpha,\beta}(R_T)\notag\\
\mathrm{s.t.}\quad&\mathbb{E}[\xi_T X_T]\le x,~X_T\ge 0.
\end{align} 
This boundary problem is treated separately because an optimizer may
satisfy $\mathbb P(X_T=0)>0$, in which case
$\mathbb E[R_T]=-\infty$. The resulting terminal wealth is called the minimum-RVaR terminal wealth.

In what follows, we derive the quantile formulation for the
mean--risk problem for $0<\lambda<1$, and subsequently analyze the boundary case
$\lambda=0$ separately.

\subsection{Quantile Formulation and Optimal Solution under RVaR}
\label{subsec:quantile-WVaR}
 
We now derive a quantile representation of the static terminal-wealth
problem. Throughout this subsection, let $F_\xi$ and $G_\xi$ denote,
respectively, the distribution function and the right-continuous
quantile function of the state-price density $\xi_T$.

\begin{condition}
\label{ass:regularity-xi}
The distribution function $F_\xi$ of the state-price density $\xi_T$
is continuously differentiable on $(0,\infty)$, with density
\[
f_\xi(x)>0,
\qquad x>0.
\]
\end{condition}

Since $0<\xi_T<\infty$ a.s., Assumption~\ref{ass:regularity-xi}
implies that $\xi_T$ has an atomless distribution with support
$(0,\infty)$. Consequently,
\[
\operatorname{ess\,inf}\xi_T=0,
\qquad
\operatorname{ess\,sup}\xi_T=\infty,
\]
and, by the probability integral transform,
\[
F_\xi(\xi_T)\sim U(0,1).
\]
Moreover, $F_\xi$ is strictly increasing, so that
\[
F_\xi(G_\xi(u))=u,
\qquad u\in(0,1).
\]
These properties justify the inverse-function calculations used in
Section~\ref{subsec:optimal-RVaR} and Appendices~\ref{B.1}--\ref{C.1}.

Assumption~\ref{ass:regularity-xi} holds, for example, when the market
coefficients are deterministic and
\[
\int_0^T\|\theta_t\|^2\,dt>0.
\]
In this case, $\xi_T$ has a nondegenerate lognormal distribution with
a strictly positive density on $(0,\infty)$.

The objective in (\ref{3.4}) is based on the quantile function of $R_T$, so the standard convex duality method is not directly applicable. We need to change the decision variable in (\ref{3.4}) from $X_T$ to its quantile function. This allows us to recover the hidden convexity in the problem. First, we prove that the budget constraint in (\ref{3.4}) must hold with equality, and the objective function improves with higher terminal wealth levels.

The first step is to show that the budget constraint must bind at an
optimum.

\begin{lemma}
\label{lem:binding-budget} 
Suppose that $X_T^*$ is an optimal terminal wealth for the static
problem \eqref{3.4}. Then
\(
\mathbb E[\xi_TX_T^*]=x.
\)
\end{lemma}

\begin{proof}
Suppose instead that
\(
\mathbb E[\xi_TX_T^*]<x.
\)
Then there exists $c>1$ sufficiently close to one such that
\(
\mathbb E[\xi_T(cX_T^*)]\le x.
\)
Moreover,
\[
R_T(cX_T^*)
=
R_T(X_T^*)+\frac{\log c}{T}.
\]
By cash invariance of WVaR,
\[
\rho_\Phi\!\left(
R_T(X_T^*)+\frac{\log c}{T}
\right)
=
\rho_\Phi(R_T(X_T^*))
-
\frac{\log c}{T}.
\]
Hence scaling terminal wealth upward strictly improves the
mean--risk objective, contradicting optimality. Therefore,
\(
\mathbb E[\xi_TX_T^*]=x.
\)
\end{proof}
We next use the anti-monotone rearrangement principle (see, e.g. \cite{Jin3}). 

\begin{lemma}
\label{lem:antimonotone} 
Let $X_T$ be a nonnegative $\mathcal F_T$-measurable random variable
with quantile function $G_X$. Then
\[
\mathbb E\!\left[
\xi_T
G_X\!\left(1-F_\xi(\xi_T)\right)
\right]
\le
\mathbb E[\xi_TX_T].
\]
Under Assumption~\ref{ass:regularity-xi}, if the left-hand side is
finite, equality holds if and only if
\[
X_T
=
G_X\!\left(1-F_\xi(\xi_T)\right)
\qquad\text{a.s.}
\]
\end{lemma}

Lemma~\ref{lem:antimonotone} states that, among all random variables
having a prescribed marginal distribution, the anti-monotone coupling
with the state-price density minimizes the initial cost.

Because the objective depends only on the distribution of terminal
wealth, while the anti-monotone rearrangement weakly reduces its cost,
any optimizer must satisfy
\[
X_T
=
G_X\!\left(1-F_\xi(\xi_T)\right)
\qquad\text{a.s.}
\]

From Lemmas \ref{lem:binding-budget} and \ref{lem:antimonotone}, we know that if $X_T$ is an optimal solution to (\ref{3.4}), then $X_T=G_X(1-F_\xi(\xi_T))$, where $G_X$ is the quantile function of $X_T$. 
Consequently, the following quantities are well defined in
the extended-real sense, 
\begin{align*}
\mathbb E[R_T]
&=\int_{[0,1)}\frac1T\log\frac{G_X(u)}{x}\du, \\
\mathbb E[\xi_TX_T]&=\int_{[0,1)}G_X(u)G_\xi(1-u)\du, \\
\rho_\Phi(R_T)&=-\int_{[0,1)}\frac1T\log\frac{G_X(u)}{x}\,\Phi(\du),
\end{align*}
where $\Phi$ is given by (\ref{3.2}). 
Hence the static mean--RVaR problem (\ref{3.4}) can be written in quantile form as
\begin{align}\label{3.8}
\max_{G\in\mathcal{G}}\quad&\lambda\int_{[0,1)}\frac1T\log\frac{G(u)}{x}\du+(1-\lambda)\int_{[0,1)}\frac1T\log\frac{G(u)}{x}\Phi(\du)\notag\\
\mathrm{s.t.}\quad&\int_{[0,1)}G(u)G_\xi(1-u)\du=x,
\end{align}
where the decision variable $G$ is the quantile function of terminal wealth. Once the optimal solution $G^*$ to (\ref{3.8}) is obtained, the optimal solution to (\ref{3.4}) is given by $X^*_T=G^*(1-F_\xi(\xi_T))$.

We now introduce a transformation that removes the pricing-kernel
weight from the budget constraint. 
Following the setup in \cite{Wei3}, define
\begin{equation}
w(s)
:=
\frac{
\displaystyle\int_{[0,s)}G_\xi(1-u)\du
}{
\mathbb E[\xi_T]
},
\qquad
s\in[0,1].
\label{eq:w}
\end{equation}  
Since
\[
\int_0^1 G_\xi(1-u)\,du
=
\mathbb E[\xi_T]
<\infty,
\]
the function $G_\xi(1-\cdot)$ is integrable on $(0,1)$.
Hence $w$ is absolutely continuous, and therefore continuous, on
$[0,1]$.
Moreover, since $\xi_T>0$ almost surely,
$G_\xi(u)>0$ for $u\in(0,1)$, so $w$ is strictly increasing.
Moreover,
\[
w(0)=0,
\qquad
w(1)
=
\frac{
\int_0^1G_\xi(u)\du
}{
\mathbb E[\xi_T]
}
=1.
\]
Thus $w$ admits a strictly increasing inverse
\[
w^{-1}:[0,1]\to[0,1].
\]
The mapping $w$ reparametrizes the original quantile scale so that
the pricing-kernel weight in the budget constraint is absorbed into
the change of variables. Its inverse $w^{-1}$ maps the transformed
scale back to the original quantile scale.

For $G\in\mathcal G$, define
\[
H(s):=G(w^{-1}(s)),
\qquad s\in[0,1].
\]
Because $w^{-1}$ is continuous and strictly increasing, $G\in\mathcal G$ if and only
if $H\in\mathcal G$. 

Using change of variables, we obtain
\begin{align*}
\int_{[0,1)}\log G(u)\du&=\int_{[0,1)}\log H(s)\dd w^{-1}(s), \\
\int_{[0,1)}G(u)G_\xi(1-u)\du&=\mathbb E[\xi_T]\int_{[0,1)}H(s)\ds,\\
\int_{[0,1)}\log G(u)\,\Phi(\du) &=\int_{[0,1)}\log H(s)\,\dd \Phi([0,w^{-1}(s)]).
\end{align*}
After dropping terms independent of $H$, problem~ (\ref{3.4}) reduces to the
following quantile optimization problem:
\begin{align}\label{3.10}
\max_{H\in\mathcal{G}}\quad&\lambda\int_{[0,1)}\log H(s)\dd w^{-1}(s)+(1-\lambda)\int_{[0,1)}\log H(s)\dd \Phi([0,w^{-1}(s)])\notag\\
\mathrm{s.t.}\quad &\int_{[0,1)}H(s)\ds=\frac{x}{\mathbb{E}[\xi_T]}.
\end{align}
This is a concave optimization problem, so the Lagrangian method can be used. 
 
Define
\[
\mathcal L(H;\lambda,\eta)
=
\lambda
\int_{[0,1)}
\log H(s)\dd w^{-1}(s)
+
(1-\lambda)
\int_{[0,1)}
\log H(s)\,
\dd \Phi([0,w^{-1}(s)])
-
\eta
\int_{[0,1)}H(s)\ds,
\] and
\[
\varphi_\lambda(s)
=
\lambda w^{-1}(s)
+
(1-\lambda)\Phi([0,w^{-1}(s)]),
\qquad s\in[0,1].
\]
For the convex envelope formulation below, we also introduce the left-limit
function of $\varphi_\lambda$ as 
\[
\varphi_\lambda(s-)
=
\lambda w^{-1}(s)
+
(1-\lambda)\Phi([0,w^{-1}(s))),
\qquad s\in(0,1],
\]
with $\varphi_\lambda(0-):=0$.

The Lagrangian can then be written as
\[
\mathcal L(H;\lambda,\eta)
=
\int_{[0,1)}
\log H(s)\dd \varphi_\lambda(s)
-
\eta\int_{[0,1)}H(s)\ds.
\]
We therefore consider the optimization problem
\begin{align}
\max_{H\in\mathcal G}\quad
\mathcal L(H;\lambda,\eta).
\label{3.11}
\end{align}

Because of the monotonicity constraint $H\in\mathcal G$, problem
\eqref{3.11} cannot in general be solved by pointwise maximization.
Indeed, $\varphi_\lambda$ need not be absolutely continuous; moreover,
even when it admits a density, the pointwise optimizer need not be
nondecreasing and hence may fail to belong to $\mathcal G$.

To incorporate the monotonicity constraint, let $\delta_\lambda$ denote
the largest convex function dominated by $\varphi_\lambda(\cdot-)$, that
is, the convex envelope of $\varphi_\lambda(\cdot-)$. Equivalently,
\begin{equation}
\delta_\lambda(s)
=
\inf_{\substack{0\le a\le s\le b\le1\\ a<b}}
\frac{
(b-s)\varphi_\lambda(a-)
+
(s-a)\varphi_\lambda(b-)
}{
b-a
},
\qquad s\in[0,1].
\label{eq:GCM}
\end{equation}
Under the standing assumptions,
$\delta_\lambda(0)=0$, 
$\delta_\lambda(1)=1$.

The optimal solution to \eqref{3.11} can then be characterized in terms
of the right derivative of $\delta_\lambda$, as stated in the following
proposition.

\begin{proposition}
\label{pro3.5}
For $0<\lambda<1$ and a fixed Lagrange multiplier $\eta>0$, an optimal solution of the
Lagrangian problem (\ref{3.11}) is
\[
H^*(s;\lambda,\eta)
=
\frac{1}{\eta}
\delta_\lambda'(s),
\qquad
s\in[0,1),
\]
where $\delta_\lambda'$ denotes the right derivative of
$\delta_\lambda$. The boundary case $\lambda=0$ is treated separately in
Section~\ref{subsec:optimal-RVaR}.
\end{proposition}
The proof is provided in Appendix \ref{A.1}.

Since $\delta_\lambda$ is convex, its right derivative exists and is
nondecreasing. Hence $H^*$ is a valid quantile function. Moreover,
\[
\int_0^1\delta_\lambda'(s)\ds
=
\delta_\lambda(1)-\delta_\lambda(0)
=
1.
\]
Therefore,
\[
\int_0^1H^*(s;\lambda,\eta)\ds
=
\frac1\eta.
\]
Imposing the budget constraint for problem \eqref{3.11} gives
\[
\eta
=
\frac{\mathbb E[\xi_T]}{x}.
\]

Returning to the original quantile scale,
\[
G^*(u)
=
H^*(w(u)). 
\]
Hence, we obtain Proposition \ref{pro3.6}.
\begin{proposition}\label{pro3.6}
For $0<\lambda<1$, the efficient terminal wealth under WVaR, i.e., the optimal solution to (\ref{3.8}), is
\begin{align}
X^*_{T,\lambda}=\frac{x}{\mathbb{E}[\xi_T]}\delta_\lambda'(w(1-F_\xi(\xi_T))),
\end{align}
and the corresponding logarithmic return is 
\[
R_{T,\lambda}^*
=
\frac1T
\log
\left[
\frac{
\delta_\lambda'
\left(
w(1-F_\xi(\xi_T))
\right)
}{
\mathbb E[\xi_T]
}
\right].
\]
\end{proposition}
 
\subsection{Optimal Solution under RVaR}
 \label{subsec:optimal-RVaR} 

We now specialize the general WVaR solution developed in
Section~\ref{subsec:quantile-WVaR} to RVaR. The key step is to
characterize explicitly the convex envelope of the RVaR-specific
transformed weighting function $\varphi_\lambda$. Since RVaR assigns
weight to the quantile interval $[\alpha,\beta]$, the function
$\varphi_\lambda$ has three pieces, with changes in slope occurring
at $w(\alpha)$ and $w(\beta)$.

For convenience, define
\[
A_\lambda
:=
\lambda+\frac{1-\lambda}{\beta-\alpha}.
\]

Since the RVaR weighting measure is atomless,
\[
\Phi_{\alpha,\beta}(\{u\})=0,
\qquad u\in[0,1].
\]
Hence
\[
\varphi_\lambda(s)-\varphi_\lambda(s-)
=
(1-\lambda)
\Phi_{\alpha,\beta}\bigl(\{w^{-1}(s)\}\bigr)
=
0,
\qquad s\in[0,1].
\]
Thus $\varphi_\lambda$ has no jumps, and the convex envelope defined
in \eqref{eq:GCM} may be constructed directly from
$\varphi_\lambda$. Accordingly, we write $\varphi_\lambda(s)$ rather
than $\varphi_\lambda(s-)$ throughout the RVaR analysis.

For RVaR,
\begin{equation}
\varphi_\lambda(s)
=
\begin{cases}
\lambda w^{-1}(s),
& s\in[0,w(\alpha)],\\[1mm]
A_\lambda w^{-1}(s)
-\dfrac{(1-\lambda)\alpha}{\beta-\alpha},
& s\in(w(\alpha),w(\beta)],\\[2mm]
\lambda w^{-1}(s)+1-\lambda,
& s\in(w(\beta),1].
\end{cases}
\label{eq:phi-RVaR}
\end{equation}
The three expressions agree at the two junction points. Indeed,
\[
\lambda w^{-1}(w(\alpha))
=
\lambda\alpha,
\]
while
\[
A_\lambda\alpha
-\frac{(1-\lambda)\alpha}{\beta-\alpha}
=
\lambda\alpha,
\]
and similarly,
\[
A_\lambda\beta
-\frac{(1-\lambda)\alpha}{\beta-\alpha}
=
\lambda\beta+1-\lambda.
\]
Therefore, $\varphi_\lambda$ is continuous on $[0,1]$.

Since $w$ is strictly increasing and concave, its inverse $w^{-1}$
is strictly increasing and convex. It follows that
$\varphi_\lambda$ is nondecreasing on $[0,1]$ and convex on each of
the three intervals separated by $w(\alpha)$ and $w(\beta)$.

At points where $w^{-1}$ is differentiable,
\[
\varphi_\lambda'(s)
=
\begin{cases}
\lambda (w^{-1})'(s),
& s\in(0,w(\alpha)),\\[1mm]
A_\lambda (w^{-1})'(s),
& s\in(w(\alpha),w(\beta)),\\[1mm]
\lambda (w^{-1})'(s),
& s\in(w(\beta),1).
\end{cases}
\]
For $0\leq\lambda<1$,
\[
A_\lambda-\lambda
=
\frac{1-\lambda}{\beta-\alpha}>0.
\]
Consequently, the slope jumps upward at $w(\alpha)$ and downward at
$w(\beta)$. More precisely, whenever the relevant one-sided
derivatives exist,
\[
\varphi_\lambda'(w(\alpha)+)
>
\varphi_\lambda'(w(\alpha)-),
\]
whereas
\[
\varphi_\lambda'(w(\beta)+)
<
\varphi_\lambda'(w(\beta)-).
\]
The upward jump at $w(\alpha)$ is consistent with convexity, whereas
the downward jump at $w(\beta)$ may violate global convexity.
Accordingly, the convex envelope construction is needed to bridge
the nonconvex region generated by the latter slope reversal.

Moreover,
\[
(w^{-1})'(s)
=
\frac{\mathbb E[\xi_T]}
{G_\xi(1-w^{-1}(s))}.
\]

For $s\in[w(\alpha),w(\beta)]$, define
\[
y(s)
:=
w\left(
1-F_\xi\left(
\frac{\lambda}{A_\lambda}
G_\xi(1-w^{-1}(s))
\right)
\right),
\]
where
\[
\frac{\lambda}{A_\lambda}
=
\frac{\lambda(\beta-\alpha)}
{1-\lambda+\lambda(\beta-\alpha)}.
\]
The point $y(s)$ is chosen so that the slope of the middle branch at
$s$ matches the slope of the right branch at $y(s)$:
\[
\varphi_\lambda'(s)
=
\varphi_\lambda'(y(s)).
\]
For the boundary case $\lambda=0$, we set $y(s):=1$.

Define the secant slope
\[
k(s)
:=
\frac{
\varphi_\lambda(y(s))-\varphi_\lambda(s)
}{
y(s)-s
},
\qquad
s\in(w(\alpha),w(\beta)),
\]
and let
\[
k_\alpha
:=
\lim_{s\downarrow w(\alpha)}k(s),
\]
whenever the limit is finite.

Set
\[
d_\alpha
:=
\varphi_\lambda'(w(\alpha)+),
\qquad
d_\beta
:=
\varphi_\lambda'(w(\beta)+).
\]
We distinguish the following two configurations:
\[
\text{\rm Scenario 1:}\qquad
d_\alpha\ge d_\beta
\quad\text{and}\quad
k_\alpha\le d_\alpha,
\]
and Scenario~2 otherwise.

\begin{lemma}
\label{lem3.7} 
Let $\delta_\lambda(\cdot)$ denote the convex envelope of
$\varphi_\lambda(\cdot)$.

\begin{enumerate}
\item
If $\lambda=0$ and Scenario~1 holds, then
\[
\delta_0(s)
=
\begin{cases}
0,
&
0\le s\le w(\alpha),
\\[1mm]
\dfrac{s-w(\alpha)}
{1-w(\alpha)},
&
w(\alpha)<s\le1.
\end{cases}
\]

\item
If $0<\lambda<1$ and Scenario~1 holds, then
\[
\delta_\lambda(s)
=
\begin{cases}
\varphi_\lambda(s),
&
0\le s\le w(\alpha),
\\[1mm]
\varphi_\lambda(w(\alpha))
+
\varphi_\lambda'(t_0(\lambda))
\bigl(s-w(\alpha)\bigr),
&
w(\alpha)<s\le t_0(\lambda),
\\[1mm]
\varphi_\lambda(s),
&
t_0(\lambda)<s\le1,
\end{cases}
\]
where $t_0(\lambda)>w(\beta)$ is the unique contact point satisfying
\[
\varphi_\lambda'(t_0(\lambda))
=
\frac{
\varphi_\lambda(t_0(\lambda))
-
\varphi_\lambda(w(\alpha))
}{
t_0(\lambda)-w(\alpha)
}.
\]

\item
If $\lambda=0$ and Scenario~2 holds, then
\[
\delta_0(s)
=
\begin{cases}
\varphi_0(s),
&
0\le s\le t_0,
\\[1mm]
\varphi_0(t_0)+\varphi_0'(t_0)(s-t_0),
&
t_0<s\le1,
\end{cases}
\]
where $w(\alpha)<t_0<w(\beta)$ is the unique contact point satisfying
\[
\varphi_0'(t_0)
=
\frac{1-\varphi_0(t_0)}
{1-t_0}.
\]

\item
If $0<\lambda<1$ and Scenario~2 holds, then
\[
\delta_\lambda(s)
=
\begin{cases}
\varphi_\lambda(s),
&
0\le s\le t_1(\lambda),
\\[1mm]
\varphi_\lambda(t_1(\lambda))
+
\varphi_\lambda'(t_1(\lambda))
\bigl(s-t_1(\lambda)\bigr),
&
t_1(\lambda)<s\le t_2(\lambda),
\\[1mm]
\varphi_\lambda(s),
&
t_2(\lambda)<s\le1,
\end{cases}
\]
where
\[
w(\alpha)<t_1(\lambda)<w(\beta)
<t_2(\lambda)<1
\]
are the two contact points of the common supporting line, satisfying
\[
\varphi_\lambda'(t_1(\lambda))
=
\varphi_\lambda'(t_2(\lambda))
\]
and
\[
\varphi_\lambda'(t_1(\lambda))
=
\frac{
\varphi_\lambda(t_2(\lambda))
-
\varphi_\lambda(t_1(\lambda))
}{
t_2(\lambda)-t_1(\lambda)
}.
\]
\end{enumerate}
\end{lemma}

The proof is provided in Appendix \ref{B.1}. 
\begin{figure}[H]
	\centering 
	\vspace{-0.35cm} 
	\subfigtopskip=2pt 
	\subfigbottomskip=2pt   
	\subfigcapskip=-5pt  
	\subfigure[$\lambda=0$, Scenario 1]{
		\label{q1}
		\includegraphics[width=0.4\linewidth]{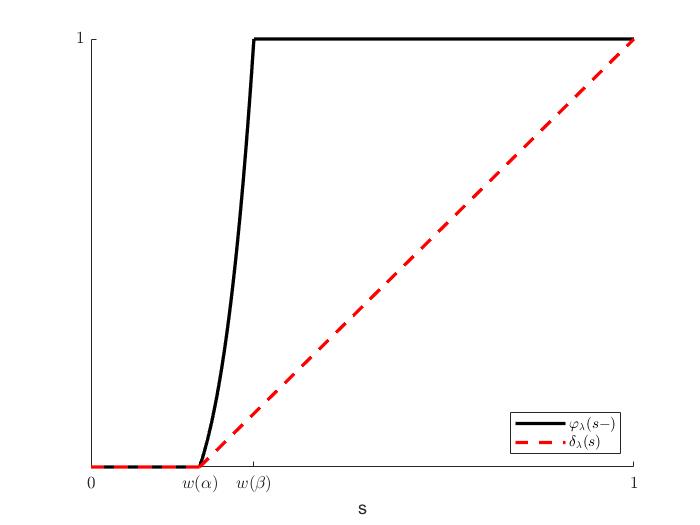}}
	\quad  
	\subfigure[$0<\lambda<1$, Scenario 1]{
		\label{q3}
		\includegraphics[width=0.4\linewidth]{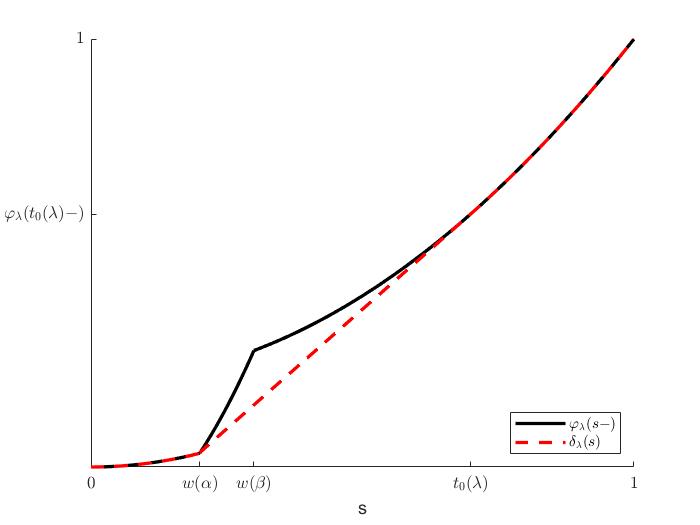}}
	\subfigure[$\lambda=0$, Scenario 2]{
		\label{q2}
		\includegraphics[width=0.4\linewidth]{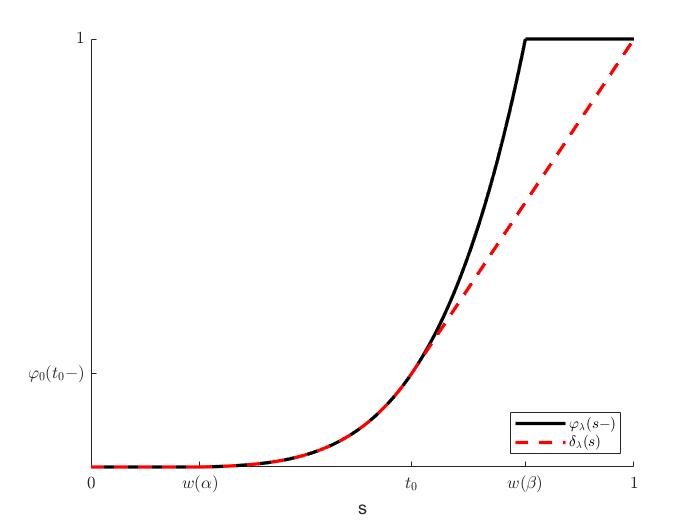}}
	\quad
	\subfigure[$0<\lambda<1$, Scenario 2]{
		\label{q4}
		\includegraphics[width=0.4\linewidth]{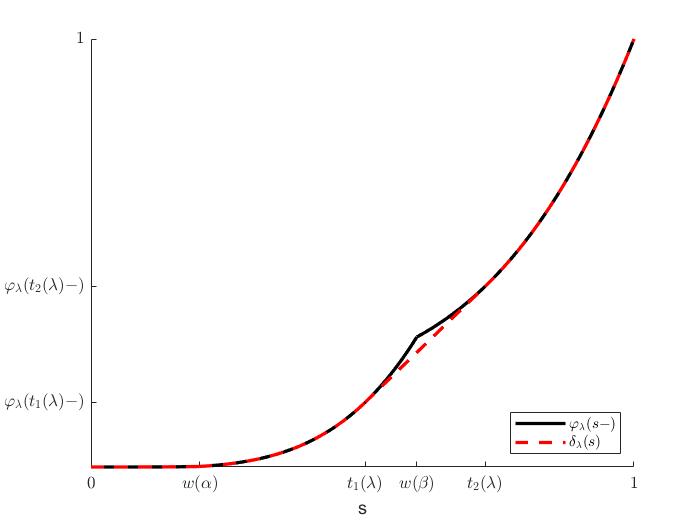}}
	\caption{Comparison of convex envelopes under different scenarios}
	\label{f1}
\end{figure}

Figure \ref{f1} provides a comparison of convex envelopes under different scenarios. After obtaining the convex envelope, the optimal solution under various RVaR scenarios can be derived.
 
\begin{proposition}[Scenario~1]  
\label{pro3.8}  
Suppose that Scenario~1 holds.

\begin{enumerate}
\item
If $\lambda=0$, the minimum-RVaR terminal wealth is
\[
X_{T,0}^{\mathrm{RVaR}}
=
\begin{cases}
0,
&
\xi_T>\xi_{1-\alpha},
\\[1mm]
X_{\mathrm{RVaR}}^{(0)},
&
0<\xi_T\le\xi_{1-\alpha},
\end{cases}
\]
where
\[
X_{\mathrm{RVaR}}^{(0)}
=
\frac{x}
{\mathbb E[\xi_T]\,[1-w(\alpha)]},
\qquad
\xi_{1-\alpha}
=
G_\xi(1-\alpha).
\]

\item
If $0<\lambda<1$, the optimal terminal wealth for the mean--RVaR
problem is
\[
X_{T,\lambda}^{\mathrm{RVaR}}
=
\begin{cases}
\dfrac{\lambda x}{\xi_T},
&
\xi_T>\xi_{1-\alpha},
\medskip\\
\widehat X_{\mathrm{RVaR}},
&
\widehat\xi_{\mathrm{RVaR}}
<\xi_T\le\xi_{1-\alpha},
\medskip\\ 
\dfrac{\lambda x}{\xi_T},
&
0<\xi_T\le\widehat\xi_{\mathrm{RVaR}},
\end{cases}
\]
where
\[
\widehat\xi_{\mathrm{RVaR}}
=
G_\xi\!\left(
1-w^{-1}(t_0(\lambda))
\right),
\qquad
\widehat X_{\mathrm{RVaR}}
=
\frac{\lambda x}
{\widehat\xi_{\mathrm{RVaR}}}.
\]
\end{enumerate}
\end{proposition}

The proof is provided in Appendix \ref{C.1}. 

Since the optimal terminal wealth is known, we can solve for the optimal wealth and portfolio strategy at time $t$ when $r,\theta$ are constant.

\begin{corollary}[Scenario~1 with constant coefficients]
\label{cor3.9}
Suppose that $r$ and $\theta$ are constants and that Scenario~1 holds.

\begin{enumerate}

\item
If $\lambda=0$, the wealth process of the minimum-RVaR portfolio is
\begin{align}
X_{t,0}^{\mathrm{RVaR}}
=
X_{\mathrm{RVaR}}^{(0)}
e^{-r(T-t)}
N\!\left(d_2(t,\xi_t,\xi_{1-\alpha})\right). \notag
\end{align}
The corresponding optimal portfolio strategy is
\begin{align}
\pi_{t,0}^{\mathrm{RVaR}}
=
\frac{
X_{\mathrm{RVaR}}^{(0)}e^{-r(T-t)}
\nu\!\left(d_2(t,\xi_t,\xi_{1-\alpha})\right)
}{
|\theta|\sqrt{T-t}
}
(\sigma_t^\top)^{-1}\theta. \notag
\end{align}

\item
If $0<\lambda<1$, the wealth process of the mean--RVaR portfolio is
\begin{align}
X_{t,\lambda}^{\mathrm{RVaR}}
={}&
\frac{\lambda x}{\xi_t}
\left[
N\!\left(-d_1(t,\xi_t,\xi_{1-\alpha})\right)
+
N\!\left(d_1(t,\xi_t,\widehat\xi_{\mathrm{RVaR}})\right)
\right]
\notag\\
&+
\widehat X_{\mathrm{RVaR}}e^{-r(T-t)}
\left[
N\!\left(d_2(t,\xi_t,\xi_{1-\alpha})\right)
-
N\!\left(d_2(t,\xi_t,\widehat\xi_{\mathrm{RVaR}})\right)
\right]. \notag
\end{align}
The corresponding optimal portfolio strategy is
\begin{align}
\pi_{t,\lambda}^{\mathrm{RVaR}}
={}&
\Bigg\{
\frac{\lambda x}{\xi_t}
\left[
N\!\left(-d_1(t,\xi_t,\xi_{1-\alpha})\right)
+
N\!\left(d_1(t,\xi_t,\widehat\xi_{\mathrm{RVaR}})\right)
\right]
\notag\\
&\quad+
\frac{
e^{-r(T-t)}
\nu\!\left(d_2(t,\xi_t,\xi_{1-\alpha})\right)
}{
|\theta|\sqrt{T-t}
}
\left(
\widehat X_{\mathrm{RVaR}}
-
\frac{\lambda x}{\xi_{1-\alpha}}
\right)
\Bigg\}
(\sigma_t^\top)^{-1}\theta.\notag
\end{align}

\end{enumerate}

Here,
\begin{align}
d_1(t,\xi_t,y)
&=
\frac{
\ln(y/\xi_t)
+
\left(r+\frac{1}{2}|\theta|^2\right)(T-t)
}{
|\theta|\sqrt{T-t}
},
\notag\\
d_2(t,\xi_t,y)
&=
d_1(t,\xi_t,y)-|\theta|\sqrt{T-t}, \notag
\end{align}
and, $N(\cdot)$ is the standard normal cumulative distribution function and $\nu(\cdot)$ is the standard normal probability density function.
\end{corollary}

The proof is provided in Appendix \ref{D.1}.

\begin{proposition}[Scenario~2]
\label{pro3.10} 
Suppose that Scenario~2 holds.

\begin{enumerate}
\item
If $\lambda=0$, the minimum-RVaR terminal wealth is
\[
X_{T,0}^{\mathrm{RVaR}}
=
\begin{cases}
0,
&
\xi_T>\xi_{1-\alpha},
\medskip\\ 
\dfrac{x}{(\beta-\alpha)\xi_T},
&
\widetilde\xi_{\mathrm{RVaR}}
<\xi_T\le\xi_{1-\alpha},
\medskip\\ 
\widetilde X_{\mathrm{RVaR}},
&
0<\xi_T\le
\widetilde\xi_{\mathrm{RVaR}},
\end{cases}
\]
where
\[
\widetilde\xi_{\mathrm{RVaR}}
=
G_\xi\!\left(1-w^{-1}(t_0)\right),
\qquad
\widetilde X_{\mathrm{RVaR}}
=
\frac{x}
{(\beta-\alpha)\widetilde\xi_{\mathrm{RVaR}}}.
\]

\item
If $0<\lambda<1$, define
\[
\bar\xi_{\mathrm{RVaR}}
=
G_\xi\!\left(
1-w^{-1}(t_1(\lambda))
\right)
\]
and
\[
\underline\xi_{\mathrm{RVaR}}
=
\frac{\lambda}{A_\lambda}
\bar\xi_{\mathrm{RVaR}}
=
\frac{\lambda(\beta-\alpha)}
{1-\lambda+\lambda(\beta-\alpha)}
\bar\xi_{\mathrm{RVaR}}.
\]
Then
\[
X_{T,\lambda}^{\mathrm{RVaR}}
=
\begin{cases}
\dfrac{\lambda x}{\xi_T},
&
\xi_T\ge\xi_{1-\alpha},
\medskip\\ 
A_\lambda\dfrac{x}{\xi_T},
&
\bar\xi_{\mathrm{RVaR}}
\le\xi_T<\xi_{1-\alpha},
\medskip\\ 
\underline X_{\mathrm{RVaR}},
&
\underline\xi_{\mathrm{RVaR}}
\le\xi_T<\bar\xi_{\mathrm{RVaR}},
\medskip\\ 
\dfrac{\lambda x}{\xi_T},
&
0<\xi_T<
\underline\xi_{\mathrm{RVaR}},
\end{cases}
\]
where
\[
\underline X_{\mathrm{RVaR}}
=
A_\lambda
\frac{x}{\bar\xi_{\mathrm{RVaR}}}
=
\lambda
\frac{x}{\underline\xi_{\mathrm{RVaR}}}.
\]
\end{enumerate}
\end{proposition}

The proof of Proposition \ref{pro3.10} is analogous to that of Proposition \ref{pro3.8}.
For full details, please refer to Appendix \ref{C.1}.

\begin{corollary}[Scenario~2 with constant coefficients]
\label{cor3.11}
Suppose that $r$ and $\theta$ are constants and that Scenario~2 holds.

\begin{enumerate}
\item
If $\lambda=0$, the wealth process of the minimum-RVaR portfolio is
\begin{align}
X_{t,0}^{\mathrm{RVaR}}
={}&
\frac{x}{(\beta-\alpha)\xi_t}
\left[
N\!\left(d_1(t,\xi_t,\xi_{1-\alpha})\right)
-
N\!\left(d_1(t,\xi_t,\widetilde\xi_{\mathrm{RVaR}})\right)
\right]
\notag\\
&+
\widetilde X_{\mathrm{RVaR}}e^{-r(T-t)}
N\!\left(d_2(t,\xi_t,\widetilde\xi_{\mathrm{RVaR}})\right). \notag
\end{align}
The corresponding optimal portfolio strategy is
\begin{align}
\pi_{t,0}^{\mathrm{RVaR}}
={}&
\frac{x}{(\beta-\alpha)\xi_t}
\Bigg[
N\!\left(d_1(t,\xi_t,\xi_{1-\alpha})\right)
-
N\!\left(d_1(t,\xi_t,\widetilde\xi_{\mathrm{RVaR}})\right)
\notag\\
&\qquad\qquad\quad+
\frac{
\nu\!\left(d_1(t,\xi_t,\xi_{1-\alpha})\right)
}{
|\theta|\sqrt{T-t}
}
\Bigg]
(\sigma_t^\top)^{-1}\theta. \notag
\end{align}

\item
If $0<\lambda<1$, the wealth process of the mean--RVaR portfolio is
\begin{align}
X_{t,\lambda}^{\mathrm{RVaR}}
={}&
\frac{\lambda x}{\xi_t}
N\!\left(-d_1(t,\xi_t,\xi_{1-\alpha})\right)
\notag\\
&+
\frac{A_\lambda x}{\xi_t}
\left[
N\!\left(d_1(t,\xi_t,\xi_{1-\alpha})\right)
-
N\!\left(d_1(t,\xi_t,\bar\xi_{\mathrm{RVaR}})\right)
\right]
\notag\\
&+
\underline X_{\mathrm{RVaR}}e^{-r(T-t)}
\left[
N\!\left(d_2(t,\xi_t,\bar\xi_{\mathrm{RVaR}})\right)
-
N\!\left(d_2(t,\xi_t,\underline\xi_{\mathrm{RVaR}})\right)
\right]
\notag\\
&+
\frac{\lambda x}{\xi_t}
N\!\left(d_1(t,\xi_t,\underline\xi_{\mathrm{RVaR}})\right). \notag
\end{align}
The corresponding optimal portfolio strategy is
\begin{align}
\pi_{t,\lambda}^{\mathrm{RVaR}}
={}&
\frac{x}{\xi_t}
\Bigg\{
\lambda
N\!\left(-d_1(t,\xi_t,\xi_{1-\alpha})\right)
\notag\\
&\quad\quad+
A_\lambda
\left[
N\!\left(d_1(t,\xi_t,\xi_{1-\alpha})\right)
-
N\!\left(d_1(t,\xi_t,\bar\xi_{\mathrm{RVaR}})\right)
\right]
\notag\\
&\quad\quad+
\lambda
N\!\left(d_1(t,\xi_t,\underline\xi_{\mathrm{RVaR}})\right)
+
(A_\lambda-\lambda)
\frac{
\nu\!\left(d_1(t,\xi_t,\xi_{1-\alpha})\right)
}{
|\theta|\sqrt{T-t}
}
\Bigg\}
(\sigma_t^\top)^{-1}\theta. \notag
\end{align}

\end{enumerate}
\end{corollary}
The proof of Corollary \ref{cor3.11} follows the same conditional-pricing
argument as that of Corollary \ref{cor3.9}. Full details are provided in
Appendix \ref{D.1}.

\subsection{Limiting Cases: VaR and ES}
\label{subsec:VaR-ES-limits}

We now examine the two limiting cases of RVaR and verify that the
optimal terminal-wealth structures derived in Section~\ref{subsec:optimal-RVaR} reduce to
the corresponding VaR and ES solutions.

\subsubsection{The VaR limit}

We first consider the limiting case in which the RVaR interval
collapses to a single quantile level. Let $\alpha\in(0,1)$ be fixed.
Under the right-continuous quantile convention adopted in
Section~\ref{sec:pre1}, the right continuity of $G_X$ implies
\[
\lim_{\beta\downarrow\alpha}
\frac{1}{\beta-\alpha}
\int_\alpha^\beta G_X(u)\,du
=
G_X(\alpha).
\]
Hence,
\[
\lim_{\beta\downarrow\alpha}
\operatorname{RVaR}_{\alpha,\beta}(X)
=
\operatorname{VaR}_{\alpha}(X).
\]

We next examine how the corresponding convex-envelope geometry
behaves as $\beta\downarrow\alpha$. For
$0\leq\lambda<1$, recall that
\[
A_\lambda
=
\lambda+\frac{1-\lambda}{\beta-\alpha}.
\]
Thus,
\[
A_\lambda\longrightarrow\infty
\qquad
\text{as }\beta\downarrow\alpha.
\]
Since
\[
(w^{-1})'(w(\alpha)+)
=
\frac{\mathbb E[\xi_T]}
{G_\xi(1-\alpha)}
\in(0,\infty),
\]
we have
\[
d_\alpha
=
\varphi_\lambda'(w(\alpha)+)
=
A_\lambda (w^{-1})'(w(\alpha)+)
\longrightarrow\infty.
\]
On the other hand,
\[
d_\beta
=
\varphi_\lambda'(w(\beta)+)
=
\lambda (w^{-1})'(w(\beta)+)
\longrightarrow
\lambda (w^{-1})'(w(\alpha)+)
<\infty.
\]

To verify the second condition defining Scenario~1, let
\[
y_\alpha
:=
\lim_{s\downarrow w(\alpha)}y(s).
\]
Since
\[
\frac{\lambda}{A_\lambda}\longrightarrow0,
\]
the regularity of $F_\xi$ implies
\[
y_\alpha
=
w\left(
1-
F_\xi\left(
\frac{\lambda}{A_\lambda}
G_\xi(1-\alpha)
\right)
\right)
\longrightarrow
w(1)=1.
\]
Therefore, for $\beta>\alpha$ sufficiently close to $\alpha$,
$y_\alpha>w(\beta)$, and
\[
k_\alpha
=
\frac{
\lambda w^{-1}(y_\alpha)+1-\lambda-\lambda\alpha
}{
y_\alpha-w(\alpha)
}.
\]
Consequently,
\[
k_\alpha
\longrightarrow
\frac{1-\lambda\alpha}{1-w(\alpha)}
<\infty.
\]
Hence, for all $\beta>\alpha$ sufficiently close to $\alpha$,
\[
d_\alpha\geq d_\beta,
\qquad
k_\alpha\leq d_\alpha.
\]
Thus, Scenario~1 applies in a neighborhood of the VaR limit.

For $0<\lambda<1$, the Scenario~1 contact point
$t_0(\lambda)>w(\beta)$ satisfies
\[
\varphi_\lambda'(t_0(\lambda))
=
\frac{
\varphi_\lambda(t_0(\lambda))
-
\varphi_\lambda(w(\alpha)-)
}{
t_0(\lambda)-w(\alpha)
}.
\]
Since $t_0(\lambda)>w(\beta)$,
\[
\varphi_\lambda(t_0(\lambda))
=
\lambda w^{-1}(t_0(\lambda))+1-\lambda,
\]
\[
\varphi_\lambda'(t_0(\lambda))
=
\lambda(w^{-1})'(t_0(\lambda)),
\]
and
\[
\varphi_\lambda(w(\alpha))=\lambda\alpha.
\]
Hence $t_0(\lambda)$ is characterized by
\begin{equation}
\lambda(w^{-1})'(t_0(\lambda))
=
\frac{
\lambda w^{-1}(t_0(\lambda))
+1-\lambda-\lambda\alpha
}{
t_0(\lambda)-w(\alpha)
}.
\label{eq:VaR-contact-point}
\end{equation}
Importantly, \eqref{eq:VaR-contact-point} does not involve
$\beta$. By uniqueness of the contact point, $t_0(\lambda)$ is
therefore independent of $\beta$ whenever Scenario~1 applies.

It follows that, once Scenario~1 applies, the resulting
convex-envelope geometry and optimal terminal-wealth structure
coincide with those of the corresponding VaR problem studied by
\cite{Wei3}. In particular, the case $\lambda=0$
recovers their minimum-VaR solution, whereas for
$0<\lambda<1$ the contact point satisfies the same tangency
condition as in their mean--VaR problem.

\begin{proposition}[VaR limit]
\label{prop:VaR-limit}
Let $\alpha\in(0,1)$ be fixed and let $0\leq\lambda<1$.
For all $\beta>\alpha$ sufficiently close to $\alpha$, the
RVaR-optimal terminal wealth coincides with the corresponding
VaR-optimal terminal wealth. Consequently, as
$\beta\downarrow\alpha$,
\[
X_{T,\lambda}^{\mathrm{RVaR}}
\longrightarrow
X_{T,\lambda}^{\mathrm{VaR}}
\qquad\text{a.s..}
\]

\begin{enumerate}
\item
If $\lambda=0$, then 
\[
X_{T,0}^{\mathrm{VaR}}
=
\begin{cases}
0,
&
\xi_T>\xi_{1-\alpha},
\\[1mm]
X_{\mathrm{VaR}}^{(0)},
&
0<\xi_T\leq\xi_{1-\alpha},
\end{cases}
\]
where
\[
X_{\mathrm{VaR}}^{(0)}
=
\frac{x}
{\mathbb E[\xi_T]\,[1-w(\alpha)]},
\qquad
\xi_{1-\alpha}
=
G_\xi(1-\alpha).
\]

\item
If $0<\lambda<1$, then
\[
X_{T,\lambda}^{\mathrm{VaR}}
=
\begin{cases}
\dfrac{\lambda x}{\xi_T},
&
\xi_T>\xi_{1-\alpha},
\medskip\\
\widehat X_{\mathrm{VaR}},
&
\widehat\xi_{\mathrm{VaR}}
<\xi_T\leq\xi_{1-\alpha},
\medskip\\
\dfrac{\lambda x}{\xi_T},
&
0<\xi_T\leq\widehat\xi_{\mathrm{VaR}},
\end{cases}
\]
where
\[
\widehat\xi_{\mathrm{VaR}}
=
G_\xi\!\left(
1-w^{-1}(t_0(\lambda))
\right),
\qquad
\widehat X_{\mathrm{VaR}}
=
\frac{\lambda x}
{\widehat\xi_{\mathrm{VaR}}},
\]
and $t_0(\lambda)$ is the unique solution of
\eqref{eq:VaR-contact-point}.
\end{enumerate}
\end{proposition}

\subsubsection{The ES case}

Setting $\alpha=0$ gives
\[
\operatorname{RVaR}_{0,\beta}(X)
=
\operatorname{ES}_{\beta}(X).
\]
Moreover,
\[
w(\alpha)=w(0)=0.
\]
By Assumption~\ref{ass:regularity-xi},
\[
G_\xi(1)=+\infty,
\]
and therefore
\[
(w^{-1})'(0+)
=
\lim_{s\downarrow0}
\frac{\mathbb E[\xi_T]}
{G_\xi(1-w^{-1}(s))}
=
0.
\]
Hence the ES case falls into the geometry corresponding to
Scenario~2.

Define
\[
A_{\lambda}^{\mathrm{ES}}
:=
\lambda+\frac{1-\lambda}{\beta}
=
\frac{1-\lambda+\beta\lambda}{\beta}.
\]
It equals to $A_\lambda$ with $\alpha=0$. 

\begin{proposition}[ES Limit]
\label{prop:ES-limit}
When $\alpha=0$, the optimal terminal wealth is given as follows.

\begin{enumerate}
\item
If $\lambda=0$, then 
\[
X_{T,0}^{\mathrm{ES}}
=
\begin{cases}
\dfrac{x}{\beta\xi_T},
&
\xi_T>\widetilde\xi_{\mathrm{ES}},
\bigskip\\ 
\widetilde X_{\mathrm{ES}},
&
0<\xi_T\le\widetilde\xi_{\mathrm{ES}},
\end{cases}
\]
where
\[
\widetilde\xi_{\mathrm{ES}}
=
G_\xi\!\left(
1-w^{-1}(t_0)
\right),
\qquad
\widetilde X_{\mathrm{ES}}
=
\frac{x}
{\beta\widetilde\xi_{\mathrm{ES}}}.
\]

\item
If $0<\lambda<1$, define
\[
\bar\xi_{\mathrm{ES}}
=
G_\xi\!\left(
1-w^{-1}(t_1(\lambda))
\right),
\]
and
\[
\underline\xi_{\mathrm{ES}}
=
G_\xi\!\left(
1-w^{-1}(t_2(\lambda))
\right).
\]
Then
\[
\bar\xi_{\mathrm{ES}}
>
\underline\xi_{\mathrm{ES}},
\]
and
\[
X_{T,\lambda}^{\mathrm{ES}}
=
\begin{cases}
A_{\lambda}^{\mathrm{ES}}
\dfrac{x}{\xi_T},
&
\xi_T>\bar\xi_{\mathrm{ES}},
\\[3mm]
X_{\mathrm{ES}},
&
\underline\xi_{\mathrm{ES}}
<\xi_T\le\bar\xi_{\mathrm{ES}},
\\[2mm]
\dfrac{\lambda x}{\xi_T},
&
0<\xi_T\le\underline\xi_{\mathrm{ES}},
\end{cases}
\]
where
\[
X_{\mathrm{ES}}
=
A_{\lambda}^{\mathrm{ES}}
\frac{x}{\bar\xi_{\mathrm{ES}}}
=
\lambda
\frac{x}{\underline\xi_{\mathrm{ES}}}.
\]
\end{enumerate}
\end{proposition}

The ES formulas are obtained directly from the Scenario~2 RVaR
solution by setting $\alpha=0$. In particular,
\[
\xi_{1-\alpha}
=
G_\xi(1)
=
+\infty,
\]
so the zero-terminal-wealth region appearing in the general RVaR
minimum-risk solution disappears.

\subsubsection{The Case with Constant Coefficients}
Assume that $r$ and $\theta$ are constants in this section.  
The corresponding wealth processes and portfolio strategies can be
obtained by conditional pricing and It\^o's formula.

For the minimum-VaR portfolio,
\[
X_{t,0}^{\mathrm{VaR}}
=
X_{\mathrm{VaR}}^{(0)}
e^{-r(T-t)}
N\!\left(
d_2(t,\xi_t,\xi_{1-\alpha})
\right),
\]
and
\[
\pi_{t,0}^{\mathrm{VaR}}
=
\frac{
X_{\mathrm{VaR}}^{(0)}
e^{-r(T-t)}
\nu\!\left(
d_2(t,\xi_t,\xi_{1-\alpha})
\right)
}{
\|\theta\|\sqrt{T-t}
}
(\sigma_t^\top)^{-1}\theta.
\]

For the minimum-ES portfolio,
\[
X_{t,0}^{\mathrm{ES}}
=
\frac{x}{\beta\xi_t}
N\!\left(
-d_1(t,\xi_t,\widetilde\xi_{\mathrm{ES}})
\right)
+
\widetilde X_{\mathrm{ES}}
e^{-r(T-t)}
N\!\left(
d_2(t,\xi_t,\widetilde\xi_{\mathrm{ES}})
\right),
\]
and
\[
\pi_{t,0}^{\mathrm{ES}}
=
\frac{x}{\beta\xi_t}
N\!\left(
-d_1(t,\xi_t,\widetilde\xi_{\mathrm{ES}})
\right)
(\sigma_t^\top)^{-1}\theta.
\]

For $0<\lambda<1$, the mean-VaR wealth process is
\begin{align*}
X_{t,\lambda}^{\mathrm{VaR}}
=&
\frac{\lambda x}{\xi_t}
\left[
N\!\left(
-d_1(t,\xi_t,\xi_{1-\alpha})
\right)
+
N\!\left(
d_1(t,\xi_t,\widehat\xi_{\mathrm{VaR}})
\right)
\right] \\
&\qquad
+
\widehat X_{\mathrm{VaR}}
e^{-r(T-t)}
\left[
N\!\left(
d_2(t,\xi_t,\xi_{1-\alpha})
\right)
-
N\!\left(
d_2(t,\xi_t,\widehat\xi_{\mathrm{VaR}})
\right)
\right],
\end{align*}
and its optimal portfolio strategy is
\begin{align*}
\pi_{t,\lambda}^{\mathrm{VaR}}
=&
\frac{\lambda x}{\xi_t}
\left[
N\!\left(
-d_1(t,\xi_t,\xi_{1-\alpha})
\right)
+
N\!\left(
d_1(t,\xi_t,\widehat\xi_{\mathrm{VaR}})
\right)
\right]
(\sigma_t^\top)^{-1}\theta \\
&\qquad
+
\frac{
e^{-r(T-t)}
\nu\!\left(
d_2(t,\xi_t,\xi_{1-\alpha})
\right)
}{
\|\theta\|\sqrt{T-t}
}
\left(
\widehat X_{\mathrm{VaR}}
-
\frac{\lambda x}{\xi_{1-\alpha}}
\right)
(\sigma_t^\top)^{-1}\theta.
\end{align*}
Finally, the mean-ES wealth process is
\begin{align*}
X_{t,\lambda}^{\mathrm{ES}}
=&~
A_{\lambda}^{\mathrm{ES}}
\frac{x}{\xi_t}
N\!\left(
-d_1(t,\xi_t,\bar\xi_{\mathrm{ES}})
\right) \\
&~+
X_{\mathrm{ES}}
e^{-r(T-t)}
\left[
N\!\left(
d_2(t,\xi_t,\bar\xi_{\mathrm{ES}})
\right)
-
N\!\left(
d_2(t,\xi_t,\underline\xi_{\mathrm{ES}})
\right)
\right]
+
\frac{\lambda x}{\xi_t}
N\!\left(
d_1(t,\xi_t,\underline\xi_{\mathrm{ES}})
\right),
\end{align*}

and the corresponding portfolio strategy is
\[
\pi_{t,\lambda}^{\mathrm{ES}}
=
\frac{x}{\xi_t}
\left[
A_{\lambda}^{\mathrm{ES}}
N\!\left(
-d_1(t,\xi_t,\bar\xi_{\mathrm{ES}})
\right)
+
\lambda
N\!\left(
d_1(t,\xi_t,\underline\xi_{\mathrm{ES}})
\right)
\right]
(\sigma_t^\top)^{-1}\theta.
\]

\section{Numerical Analysis}\label{sec:numerical}

To illustrate the theoretical results and examine the portfolio implications
of RVaR, we consider a standard Black--Scholes market. The initial wealth is
$x=1$ and the investment horizon is $T=1$. The market parameters are set to
\[
r=0.05,\qquad \sigma=0.15,\qquad \theta=0.4.
\]
Under these parameters, the Growth Optimal Portfolio (GOP) allocates
approximately
\[
\sigma^{-1}\theta =2.6667
\]
times initial wealth to the risky asset at time $0$. We use the initial risky
asset position to illustrate how the RVaR parameters affect optimal portfolio
choice.

\subsection{Terminal Wealth Structure}

We first illustrate the optimal terminal wealth $X_T$ as a function of the
state-price density $\xi_T$ under the two scenarios characterized in
Section~\ref{sec:strategy_RVaR}.

\begin{figure}[H]
\centering
\vspace{-0.35cm}
\subfigtopskip=2pt
\subfigbottomskip=2pt
\subfigcapskip=-5pt
\subfigure[$\lambda=0$, Scenario 1]{
\label{f11}
\includegraphics[width=0.45\linewidth]{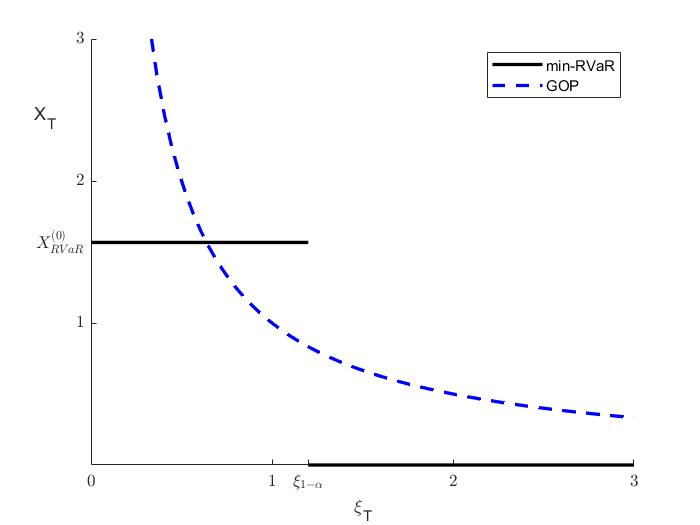}}
\quad
\subfigure[$0<\lambda<1$, Scenario 1]{
\label{f12}
\includegraphics[width=0.45\linewidth]{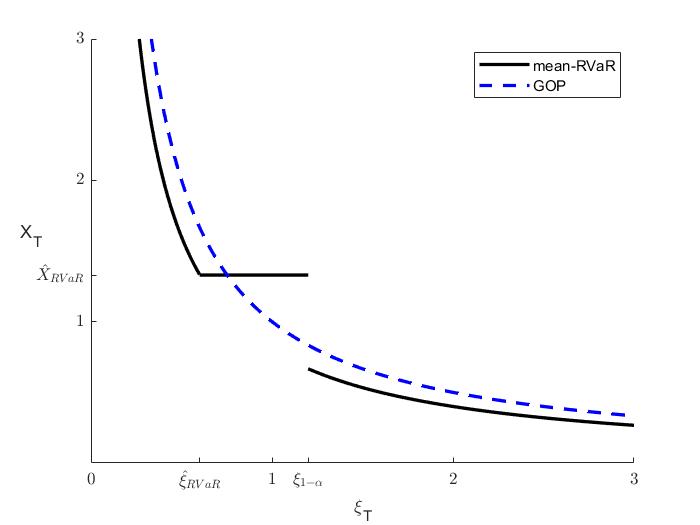}}
\caption{Optimal terminal wealth under RVaR in Scenario~1.}
\label{f2}
\end{figure}

The left panel of Figure~\ref{f2} illustrates the minimum-RVaR terminal wealth
under Scenario~1. When $\lambda=0$, the optimal payoff has a
digital-option-type structure:
\[
X_{T,0}^{\mathrm{RVaR}}
=
\begin{cases}
0, & \xi_T>\xi_{1-\alpha},\medskip\\
X_{\mathrm{RVaR}}^{(0)}, & 0<\xi_T\le\xi_{1-\alpha}.
\end{cases}
\]
Since a smaller state-price density corresponds to a more favorable market
state, the investor receives the constant payoff
$X_{\mathrm{RVaR}}^{(0)}$ in sufficiently favorable states and zero wealth in
sufficiently adverse states.

The right panel depicts the mean--RVaR solution for $0<\lambda<1$. The state
space is divided into three regions. In favorable states
$\xi_T\le\widehat{\xi}_{\mathrm{RVaR}}$ and sufficiently adverse states
$\xi_T>\xi_{1-\alpha}$, terminal wealth is proportional to the GOP payoff,
\[
X_{T,\lambda}^{\mathrm{RVaR}}=\frac{\lambda x}{\xi_T}.
\]
In the intermediate region
$\widehat{\xi}_{\mathrm{RVaR}}<\xi_T\le\xi_{1-\alpha}$, terminal wealth is
constant and equal to $\widehat X_{\mathrm{RVaR}}$. Thus, the convex-envelope
construction replaces part of the GOP-type payoff by a flat segment.

Under Scenario~1, the resulting payoff structure is closely related to the
corresponding VaR solution. In particular, the same type of digital payoff
arises for $\lambda=0$, while for $0<\lambda<1$ both problems generate a
GOP--constant--GOP structure. This structural similarity follows from the
convex-envelope geometry established in Section~\ref{sec:strategy_RVaR} rather than from a general
dominance relation between RVaR and VaR.

\begin{figure}[H]
\centering
\vspace{-0.35cm}
\subfigtopskip=2pt
\subfigbottomskip=2pt
\subfigcapskip=-5pt
\subfigure[$\lambda=0$, Scenario 2]{
\label{f21}
\includegraphics[width=0.45\linewidth]{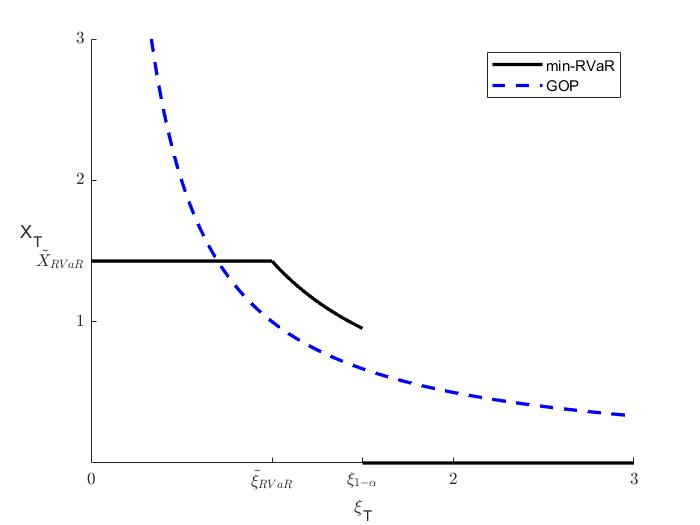}}
\quad
\subfigure[$0<\lambda<1$, Scenario 2]{
\label{f22}
\includegraphics[width=0.45\linewidth]{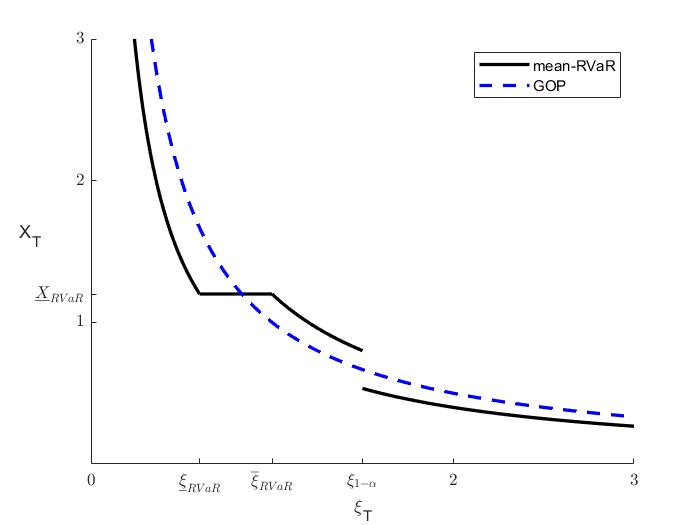}}
\caption{Optimal terminal wealth under RVaR in Scenario~2.}
\label{f3}
\end{figure}

Figure~\ref{f3} illustrates the richer payoff structures arising under
Scenario~2. When $\lambda=0$, the optimal terminal wealth has three regions:
\[
X_{T,0}^{\mathrm{RVaR}}
=
\begin{cases}
0, & \xi_T>\xi_{1-\alpha}, \medskip\\
\dfrac{x}{(\beta-\alpha)\xi_T},
& \widetilde{\xi}_{\mathrm{RVaR}}<\xi_T\le\xi_{1-\alpha}, \medskip\\
\widetilde X_{\mathrm{RVaR}},
& 0<\xi_T\le\widetilde{\xi}_{\mathrm{RVaR}}.
\end{cases}
\]
Hence, terminal wealth is constant in favorable states, proportional to
$1/\xi_T$ in intermediate states, and zero in sufficiently adverse states.

When $0<\lambda<1$, the optimal payoff contains four regions. In sufficiently
favorable states
$0<\xi_T<\underline{\xi}_{\mathrm{RVaR}}$ and sufficiently adverse states
$\xi_T\ge\xi_{1-\alpha}$, the payoff equals $\lambda x/\xi_T$. In the region
$\underline{\xi}_{\mathrm{RVaR}}\le\xi_T<\bar{\xi}_{\mathrm{RVaR}}$, the
payoff is constant at $\underline X_{\mathrm{RVaR}}$, whereas in
$\bar{\xi}_{\mathrm{RVaR}}\le\xi_T<\xi_{1-\alpha}$ it is
\[
A_\lambda\frac{x}{\xi_T},
\qquad
A_\lambda=\lambda+\frac{1-\lambda}{\beta-\alpha}.
\]
This four-region structure reflects the common-supporting-line geometry of
the convex envelope in Scenario~2.

\begin{remark}
When $\lambda=0$, both Scenario~1 and Scenario~2 may assign zero terminal
wealth to sufficiently adverse states. In such states,
\[
X_T=0
\qquad\Longrightarrow\qquad
R_T=-\infty.
\]
This feature should not be interpreted as protection against extreme losses.
Rather, it arises because RVaR with $\alpha>0$ evaluates only the quantile
interval $[\alpha,\beta]$ and excludes the more extreme region
$[0,\alpha)$. Consequently, pure minimum-RVaR optimization may improve the
targeted RVaR criterion by sacrificing states outside the penalized quantile
range. This illustrates an important limitation of minimizing a localized
risk measure in isolation and motivates the inclusion of expected
logarithmic growth when $0<\lambda<1$.
\end{remark}

\begin{figure}[H]
\centering
\vspace{-0.35cm}
\subfigtopskip=2pt
\subfigbottomskip=2pt
\subfigcapskip=-5pt
\subfigure[$\lambda=0$, Scenario 2]{
\label{f31}
\includegraphics[width=0.45\linewidth]{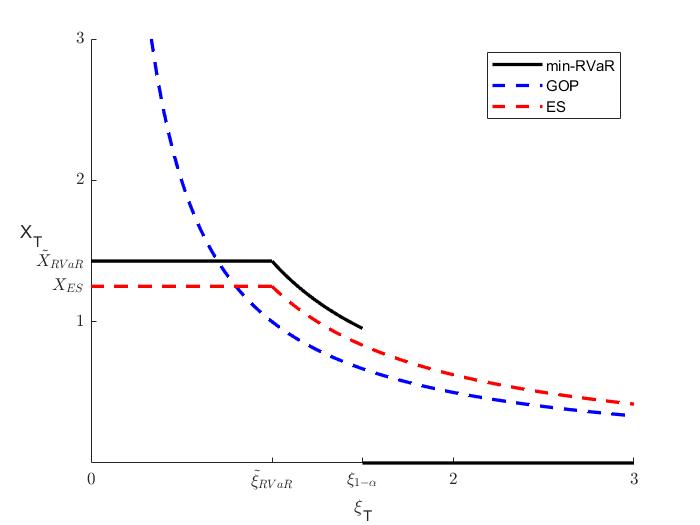}}
\quad
\subfigure[$0<\lambda<1$, Scenario 2]{
\label{f32}
\includegraphics[width=0.45\linewidth]{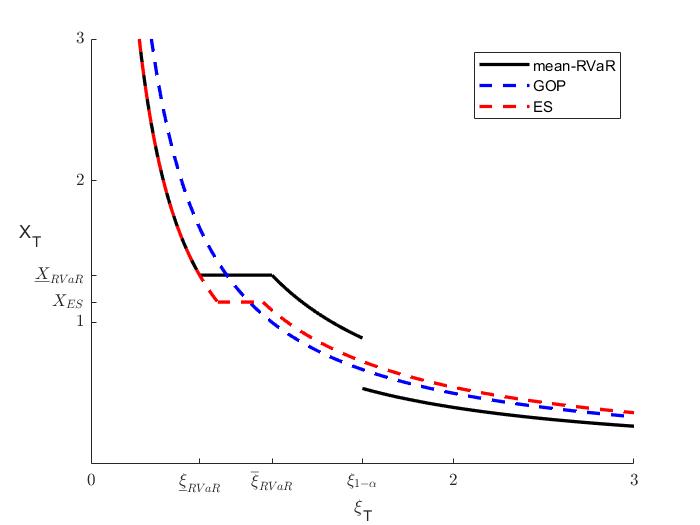}}
\caption{Comparison of optimal terminal wealth under RVaR and ES in
Scenario~2.}
\label{f4}
\end{figure}

Figure~\ref{f4} compares the RVaR and ES terminal-wealth profiles under
Scenario~2. The parameters are set to
\[
\lambda=0.8,\qquad \alpha=0.2,\qquad \beta=0.6.
\]
Because ES corresponds to the limiting case $\alpha=0$, the two criteria
penalize different portions of the lower tail. RVaR averages quantiles only
over $[\alpha,\beta]$, whereas ES averages over $[0,\beta]$. The resulting
terminal-wealth profiles therefore differ primarily in states associated
with the lower quantile region excluded by RVaR.

The comparison highlights the localized nature of RVaR. By varying
$[\alpha,\beta]$, the investor changes the portion of the quantile
distribution entering the risk criterion. This additional degree of freedom
generates terminal-wealth profiles that may differ substantially from those
under ES, particularly in sufficiently adverse states. The comparison should
therefore be interpreted as a difference in the location of risk control,
rather than as a general dominance of one risk measure over the other.

\subsection{Parameter Sensitivity and Strategy Comparison}

We next examine how the parameters $(\alpha,\beta,\lambda)$ affect the
initial risky-asset position. The interval $[\alpha,\beta]$ determines the
quantile region entering RVaR: $\alpha$ determines the lower endpoint, while
$\beta-\alpha$ determines the width of the averaging interval. As
$\beta\downarrow\alpha$, RVaR approaches VaR, whereas $\alpha=0$ yields ES.
The parameter $\lambda$ determines the relative weight assigned to expected
logarithmic growth and RVaR.

Table~\ref{table1} reports the amount invested in the risky asset at time
$0$ under different parameter combinations.

\begin{table}[h]
\centering
\caption{Initial risky-asset position under RVaR.}
\label{table1}
\begin{tabular}{*{9}{c}}
\toprule
\multirow{2}*{}&
\multirow{2}*{}&
\multirow{2}*{}&
\multicolumn{6}{c}{$\lambda$}\\
\cmidrule(lr){4-9}
&&& 0 & 0.2 & 0.4 & 0.6 & 0.8 & 1\\
\midrule
\multirow{10}*{$\alpha$}
&\multirow{2}*{0}
&$\beta=0.01$&0&0.5333&1.0667&1.5971&1.1780&2.6667\\
&&$\beta=0.5$&0.1505&0.2509&0.4366&0.8646&1.6462&2.6667\\
&\multirow{2}*{0.01}
&$\beta=0.02$&0.4275&0.3917&0.3822&0.5472&1.1688&2.6667\\
&&$\beta=0.51$&0.2939&0.4726&0.6437&1.0063&1.7176&2.6667\\
&\multirow{2}*{0.05}
&$\beta=0.06$&1.3717&1.2002&1.1305&1.1511&1.5475&2.6667\\
&&$\beta=0.55$&1.3717&1.2698&1.2435&1.4235&1.9292&2.6667\\
&\multirow{2}*{0.1}
&$\beta=0.11$&2.2236&1.8583&1.7787&1.6547&1.8452&2.6667\\
&&$\beta=0.6$&2.2236&2.0023&1.8387&1.8461&2.1463&2.6667\\
&\multirow{2}*{0.2}
&$\beta=0.21$&3.5974&2.6455&2.7909&2.4185&2.2801&2.6667\\
&&$\beta=0.7$&3.5974&3.1902&2.8233&2.5658&2.5231&2.6667\\
\bottomrule
\end{tabular}
\end{table}

Several observations follow from Table~\ref{table1}. First, the effect of the
interval width $\beta-\alpha$ is not globally monotone. For example, when
$\alpha=0$ and $\lambda=0.2$, increasing $\beta$ from $0.01$ to $0.5$
reduces the initial risky position from $0.5333$ to $0.2509$. By contrast,
for $\alpha=0.05$ and $\lambda=0.2$, increasing $\beta$ from $0.06$ to
$0.55$ raises the position from $1.2002$ to $1.2698$. Thus, the effect of
changing the RVaR interval depends jointly on $(\alpha,\beta,\lambda)$ and on
the associated convex-envelope scenario.

Second, the dependence on $\lambda$ is also generally non-monotone. For
example, for $(\alpha,\beta)=(0.2,0.21)$, the initial risky position equals
\[
3.5974,\ 2.6455,\ 2.7909,\ 2.4185,\ 2.2801,\ 2.6667
\]
as $\lambda$ increases from $0$ to $1$. Hence, although $\lambda$ increases
the weight placed on expected logarithmic growth, it should not be interpreted
as a conventional parameter that generates a globally monotone change in
portfolio aggressiveness.

Third, increasing $\alpha$ changes which extreme quantile states are excluded
from the RVaR criterion. This may permit larger risky positions because a
larger portion of the most adverse tail is left outside the targeted risk
interval. Nevertheless, Table~\ref{table1} shows that the resulting exposure
depends on the simultaneous values of $\beta$ and $\lambda$, so no universal
monotonicity statement should be inferred from $\alpha$ alone.

The cases $\alpha=0$ and $\beta\downarrow\alpha$ provide useful benchmarks.
When $\alpha=0$, RVaR coincides with ES, while a sufficiently narrow interval
approximates VaR. The numerical results therefore illustrate how the RVaR
solutions interpolate between these benchmark structures. For instance,
when $\alpha=0$, $\beta=0.5$, and $\lambda=0.2$, the initial risky position
is $0.2509$, whereas for $\alpha=0.05$, $\beta=0.55$, and $\lambda=0.2$, it
is $1.2698$. This comparison demonstrates that excluding the most extreme
quantile region can materially alter optimal risky exposure.

Overall, the numerical results confirm the structural richness predicted by
the theoretical analysis. The optimal allocation depends jointly and
nonlinearly on the location and width of the RVaR interval and on the
return--risk weight $\lambda$. RVaR therefore provides an additional degree
of freedom relative to its VaR and ES limiting cases, but this flexibility
does not imply a uniform ordering of portfolio risk, return, or leverage
across parameter configurations.
 
\section{Conclusion}\label{sec:conclude}

This paper studies continuous-time portfolio selection under RVaR when portfolio performance and risk are both
evaluated in terms of logarithmic returns. In a complete diffusion
market, we reformulate the dynamic portfolio problem as a static
quantile optimization problem and characterize the optimal terminal
wealth through the convex envelope of an RVaR-specific
transformed weighting function.

The main theoretical contribution is the explicit characterization of
the optimal payoff structure induced by RVaR. Because RVaR averages
quantiles over the interval $[\alpha,\beta]$, the transformed
weighting function has a three-region structure. Depending on its
relative slopes and supporting-line geometry, two distinct cases
arise. Under Scenario~1, the optimal terminal wealth has a VaR-type
structure. In the pure minimum-RVaR case, the payoff is
digital-option-like, whereas for $0<\lambda<1$ it consists of two
GOP-type regions separated by a constant-payoff region. Under
Scenario~2, the convex envelope involves either a tangent to
the endpoint or a common supporting line connecting two interior
contact points, producing a richer multi-region terminal-wealth
structure.

The analysis also establishes consistency with the two principal
limiting cases of RVaR. As $\beta\downarrow\alpha$, the RVaR
criterion converges to VaR under the right-continuous quantile
convention, and the corresponding optimal payoff reduces to the
VaR-type solution. When $\alpha=0$, RVaR coincides with ES, and the
Scenario~2 solution reduces to the corresponding ES terminal-wealth
structure. These limiting results provide an internal consistency
check for the general RVaR solution.

The numerical analysis illustrates that the effects of the RVaR
parameters $(\alpha,\beta,\lambda)$ on portfolio allocation are
nonlinear. In particular, neither the width $\beta-\alpha$ nor the
return--risk weight $\lambda$ generates a globally monotone effect on
the initial risky-asset exposure. For some parameter configurations,
pure minimum-RVaR optimization can even produce a more highly
leveraged position than the growth-optimal portfolio. This behavior
reflects the localized nature of RVaR rather than conventional
risk-aversion comparative statics.

An important implication concerns the lower endpoint $\alpha$. When
$\alpha>0$, the extreme quantile region $[0,\alpha)$ is excluded from
the RVaR criterion. Consequently, in the pure minimum-RVaR problem
($\lambda=0$), the optimizer may sacrifice sufficiently adverse states
and assign zero terminal wealth to them. In this case,
\[
\mathbb P(X_T=0)>0
\]
and therefore
\[
\mathbb E[R_T]=-\infty.
\]
This phenomenon highlights an important limitation of optimizing a
localized tail-risk measure in isolation and provides a motivation for
jointly considering expected logarithmic growth and RVaR.

Several extensions merit further investigation. First, the economic
interpretation of the conditions separating Scenario~1 and
Scenario~2 could be developed further, ideally in terms of observable
market quantities and investor preferences rather than purely
geometric conditions on the transformed weighting function. Second,
the analysis is conducted in a complete diffusion market without
transaction costs or portfolio constraints. Extending the framework
to incomplete markets, stochastic investment opportunities, trading
constraints, and transaction costs would be important for practical
applications. Third, the numerical analysis focuses on optimal
state-contingent wealth and initial risky-asset exposure. A broader
comparison based on expected return, RVaR, VaR, ES, extreme-loss
probabilities, and other out-of-sample risk measures would provide a
more comprehensive assessment of the economic trade-offs generated by
RVaR.

\newpage
\appendix
\renewcommand{\appendixname}{Appendix~\Alph{section}}
\section{Proof of Proposition \ref{pro3.5} } \label{A.1}
 
The argument follows the convexification method of \cite{Wei3},
adapted to the notation and formulation of the present paper.

Recall that, for a fixed $0<\lambda<1$ and Lagrange multiplier
$\eta>0$, the Lagrangian problem is
\begin{equation}
\sup_{H\in\mathcal G}
\left\{
\int_{[0,1)} \log H(s)\dd \varphi_\lambda(s)
-
\eta\int_0^1 H(s)\ds
\right\}.
\label{eq:A1}
\end{equation}
Since $\delta_\lambda$ is convex on $[0,1]$,
its right derivative $\delta_\lambda'$ exists on $[0,1)$ and is
nondecreasing. Moreover,
\begin{equation}
\delta_\lambda(b)-\delta_\lambda(a)
=
\int_a^b \delta_\lambda'(s)\ds,
\qquad
0\le a<b\le1.
\label{eq:A2}
\end{equation}
In particular, because
$\delta_\lambda(0)=0$ and $\delta_\lambda(1)=1$, 
we have
\begin{equation}
\int_0^1 \delta_\lambda'(s)\ds=1.
\label{eq:A3}
\end{equation}

We next show how the convex envelope incorporates the
monotonicity constraint on $H$. Since $H\in\mathcal G$ is
nondecreasing, the function
\[
q(s):=\log H(s)
\]
is also nondecreasing wherever it is finite. By the
Lebesgue--Stieltjes integration-by-parts formula,
\[
\int_{[0,1)}q(s)\dd \varphi_\lambda(s)
=
q(1-)\varphi_\lambda(1-)
-
\int_{[0,1)}\varphi_\lambda(s-)\dd q(s),
\]
and similarly,
\[
\int_{[0,1)}q(s)\dd \delta_\lambda(s)
=
q(1-)\delta_\lambda(1)
-
\int_{[0,1)}\delta_\lambda(s)\dd q(s).
\]
Since
\[
\varphi_\lambda(1-)=\delta_\lambda(1)=1,
\qquad
\delta_\lambda(s)\le \varphi_\lambda(s-),
\]
and $\dd q$ is a nonnegative measure, it follows that
\begin{equation}
\int_{[0,1)}\log H(s)\dd \varphi_\lambda(s)
\le
\int_{[0,1)}\log H(s)\dd \delta_\lambda(s).
\label{eq:A4}
\end{equation}

Therefore, the maximization problem in \eqref{eq:A1} is reduced,
through the convexification argument, to
\begin{equation}
\sup_{H\in\mathcal G}
\left\{
\int_{[0,1)}\log H(s)\dd \delta_\lambda(s)
-
\eta\int_0^1H(s)\ds
\right\}.
\label{eq:A5}
\end{equation}
Since $\delta_\lambda$ is convex, it is absolutely continuous on
compact subintervals of $(0,1)$ and
\[
\dd\delta_\lambda(s)
=
\delta_\lambda'(s)\ds
\]
for Lebesgue-a.e.\ $s\in(0,1)$. Hence the objective in
\eqref{eq:A5} can be written as
\begin{equation}
\int_0^1
\left[
\delta_\lambda'(s)\log H(s)
-
\eta H(s)
\right]\ds.
\label{eq:A6}
\end{equation}

For each fixed $s$ such that $\delta_\lambda'(s)>0$, define
\[
\ell_s(h)
=
\delta_\lambda'(s)\log h-\eta h,
\qquad h>0.
\]
The function $\ell_s$ is strictly concave because
\[
\ell_s''(h)
=
-\frac{\delta_\lambda'(s)}{h^2}
<0.
\]
Its unique maximizer therefore satisfies
\[
\ell_s'(h)
=
\frac{\delta_\lambda'(s)}{h}-\eta
=
0,
\]
which yields
\[
h
=
\frac{\delta_\lambda'(s)}{\eta}.
\]

For the candidate optimizer
\[
H^*(s;\lambda,\eta)=\frac{1}{\eta}\delta_\lambda'(s),
\]
the inequality \eqref{eq:A4} is in fact an equality. Indeed, on every open
interval on which $\delta_\lambda(s)<\varphi_\lambda(s-)$, the convex envelope
$\delta_\lambda$ is affine, so $\delta_\lambda'$ is constant there and
$\log H^*$ is constant as well. Consequently,
\[
\int_{[0,1)}\bigl(\varphi_\lambda(s-)-\delta_\lambda(s)\bigr)\dd\;(\log H^*(s))=0,
\]
and applying the integration-by-parts formula as in \eqref{eq:A4} gives
\[
\int_{[0,1)}\log H^*(s)\dd \varphi_\lambda(s)
=
\int_{[0,1)}\log H^*(s)\dd \delta_\lambda(s).
\]
Let $\mathcal L_\delta$ denote the Lagrangian obtained by replacing
$\varphi_\lambda$ with $\delta_\lambda$. Then, for every admissible $H$,
\[
\mathcal L(H;\lambda,\eta)
\le
\mathcal L_\delta(H;\lambda,\eta)
\le
\mathcal L_\delta(H^*;\lambda,\eta)
=
\mathcal L(H^*;\lambda,\eta),
\]
which establishes the optimality of $H^*$ for the original Lagrangian
problem \eqref{3.11}.
Consequently,
\begin{equation}
H^*(s;\lambda,\eta)
=
\frac{1}{\eta}\delta_\lambda'(s),
\qquad
s\in[0,1).
\label{eq:A7}
\end{equation}

Because $\delta_\lambda'$ is nondecreasing,
$H^*(\cdot;\lambda,\eta)$ is nondecreasing and hence belongs to the
admissible quantile class $\mathcal G$.

For $0<\lambda<1$, the Lebesgue component of the objective has
strictly positive weight, and the optimizer is positive almost
everywhere on $(0,1)$. The boundary case $\lambda=0$ is not covered by the integration-by-parts
argument above (the optimizer may vanish on a nontrivial interval, so that
$\log H^*=-\infty$ there). It is treated separately in Section~\ref{subsec:optimal-RVaR} and
Appendix~\ref{C.1}.

Finally, by \eqref{eq:A3},
\[
\int_0^1 H^*(s;\lambda,\eta)\ds
=
\frac{1}{\eta}
\int_0^1\delta_\lambda'(s)\ds
=
\frac{1}{\eta}.
\]
Imposing the budget constraint
\[
\int_0^1H^*(s;\lambda,\eta)\ds
=
\frac{x}{\mathbb E[\xi_T]}
\]
gives
\[
\eta
=
\frac{\mathbb E[\xi_T]}{x}.
\]
This proves Proposition~\ref{pro3.5} .

\section{Proof of Lemma \ref{lem3.7}}\label{B.1}
\begin{enumerate}
\item[\bf{1)}]
When $\lambda=0$, we have
\begin{align}
\varphi_{0}(s)=
\begin{cases}
0,&s\in[0,w(\alpha)],\\
\dfrac{w^{-1}(s)-\alpha}{\beta-\alpha},
&s\in(w(\alpha),w(\beta)],\\
1,&s\in(w(\beta),1].
\end{cases}
\end{align}
Clearly, $\delta_0(s)$ is convex. For any convex function $\psi(s)$ dominated by
$\varphi_0(s)$, since $\varphi_0(s)=\delta_0(s)$ on $[0,w(\alpha)]$,
$\psi(s)$ is also dominated by $\delta_0(s)$ on this interval. By the
condition of Scenario 1, we have
\begin{align}
\frac{1}{1-w(\alpha)}
\le
\varphi_0'(w(\alpha)+).
\notag
\end{align}
This implies that $\delta_0(s)$ is dominated by $\varphi_0(s)$ on
$(w(\alpha),1]$. Since
$\delta_0(w(\alpha))=\varphi_0(w(\alpha))\ge\psi(w(\alpha))$ and
$\delta_0(1)=\varphi_0(1)\ge\psi(1)$, for any
$s\in(w(\alpha),1]$, we have
\begin{align}
\delta_0(s)
&=
\frac{1-s}{1-w(\alpha)}\delta_0(w(\alpha))
+
\frac{s-w(\alpha)}{1-w(\alpha)}\delta_0(1)
\notag\\
&=
\frac{1-s}{1-w(\alpha)}\varphi_0(w(\alpha))
+
\frac{s-w(\alpha)}{1-w(\alpha)}\varphi_0(1)
\notag\\
&\ge
\frac{1-s}{1-w(\alpha)}\psi(w(\alpha))
+
\frac{s-w(\alpha)}{1-w(\alpha)}\psi(1)
\ge
\psi(s).
\notag
\end{align}
Therefore, $\delta_0(s)$ is the largest convex function on $[0,1]$
dominated by $\varphi_0(s)$; hence, it is the convex envelope of
$\varphi_0(s)$.

\item[\bf{2)}]
Let
\begin{align}
f_\lambda(s)
=
\varphi_\lambda(s)
-
\varphi_\lambda(w(\alpha))
-
\varphi_\lambda'(s)(s-w(\alpha)),
\qquad
w(\beta)<s<1,
\notag
\end{align}
which is continuous in $s$. For
$w(\beta)<s_1<s_2<1$, using integration by parts,
\begin{align}
&f_\lambda(s_1)-f_\lambda(s_2)
\notag\\
={}&
\varphi_\lambda(s_1)-\varphi_\lambda(s_2)
-\varphi_\lambda'(s_1)(s_1-w(\alpha))
+\varphi_\lambda'(s_2)(s_2-w(\alpha))
\notag\\
={}&
-\int_{s_1-}^{s_2-}\dd\varphi_\lambda(z)
+\int_{s_1}^{s_2}\dd\;\bigl(z\varphi_\lambda'(z)\bigr)
-w(\alpha)
\bigl(\varphi_\lambda'(s_2)-\varphi_\lambda'(s_1)\bigr)
\notag\\
={}&
-\int_{s_1}^{s_2}\varphi_\lambda'(z)\dz
+\int_{s_1}^{s_2}z\dd \varphi_\lambda'(z)
+\int_{s_1}^{s_2}\varphi_\lambda'(z)\dz
-w(\alpha)\int_{s_1}^{s_2}\dd\varphi_\lambda'(z)
\notag\\
={}&
\int_{s_1}^{s_2}
(z-w(\alpha))\dd \varphi_\lambda'(z)
>0.
\notag
\end{align}
Due to the strict convexity of $\varphi_\lambda(s)$ on
$(w(\beta),1]$, $f_\lambda(s)$ is therefore a strictly decreasing
function of $s$. Meanwhile,
\begin{align}
\lim_{s\downarrow w(\beta)}f_\lambda(s)
=
\lim_{s\downarrow w(\beta)}
\left[
\varphi_\lambda(s)
-\lambda\alpha
-\varphi_\lambda'(s)(s-w(\alpha))
\right].
\notag
\end{align}
By Assumption~\ref{ass:regularity-xi}, $(w^{-1})'$ is strictly increasing on
$(w(\alpha),w(\beta))$; hence, using $A_\lambda>\lambda$ (for $\lambda<1$)
and $d_\alpha=\varphi_\lambda'(w(\alpha)+)=A_\lambda(w^{-1})'(w(\alpha)+)$,
\begin{align*}
\varphi_\lambda(w(\beta))-\varphi_\lambda(w(\alpha))
&=
A_\lambda\int_{w(\alpha)}^{w(\beta)}(w^{-1})'(z)\dz
\\
&>
A_\lambda(w^{-1})'(w(\alpha)+)\bigl(w(\beta)-w(\alpha)\bigr)
\\
&=
d_\alpha\bigl(w(\beta)-w(\alpha)\bigr)
\;\ge\;
d_\beta\bigl(w(\beta)-w(\alpha)\bigr),
\end{align*}
where the last step uses $d_\alpha\ge d_\beta$ from Scenario~1. Since
$\varphi_\lambda'(w(\beta)+)=d_\beta$, we obtain
\[
\lim_{s\downarrow w(\beta)}f_\lambda(s)
=
\varphi_\lambda(w(\beta))-\varphi_\lambda(w(\alpha))
-
d_\beta\bigl(w(\beta)-w(\alpha)\bigr)
>
0.
\]
Meanwhile 
\begin{align}
\lim_{s\uparrow1}f_\lambda(s)
=
1-\lambda\alpha
-(1-w(\alpha))
\lim_{s\uparrow1}\varphi_\lambda'(s)
=
-\infty.
\notag
\end{align}
Therefore, there exists a unique
$t_0(\lambda)\in(w(\beta),1)$ such that
$f_\lambda(t_0(\lambda))=0$.

Due to the convexity of $\varphi_\lambda(s)$ on
$[0,w(\alpha)]$, $(w(\alpha),w(\beta)]$, and $(w(\beta),1]$,
together with the condition of Scenario 1, it follows that
$\delta_\lambda(s)$ is a continuous convex function dominated by
$\varphi_\lambda(s)$ on $[0,1]$. For any convex function $\psi(s)$
dominated by $\varphi_\lambda(s)$, it is clearly dominated by
$\delta_\lambda(s)$ on
$[0,w(\alpha)]\cup(t_0(\lambda),1]$. By an argument similar to that
used in part {\bf(1)}, $\psi(s)$ is also dominated by
$\delta_\lambda(s)$ on $(w(\alpha),t_0(\lambda)]$.
Therefore, $\delta_\lambda(s)$ is the largest convex function dominated
by $\varphi_\lambda(s)$; hence, it is the convex envelope of
$\varphi_\lambda(s)$.

\item[\bf{(3)}]
Let
\begin{align}
g_0(s)
=
1-\varphi_0(s)-\varphi_0'(s)(1-s),
\qquad
w(\alpha)<s<w(\beta).
\notag
\end{align}
For $w(\alpha)<s_1<s_2<w(\beta)$, using integration by parts,
we obtain
\[
g_0(s_1)-g_0(s_2)>0,
\]
so $g_0(s)$ is a strictly decreasing function of $s$. Moreover, by
the strict convexity of $\varphi_0$ on
$(w(\alpha),w(\beta))$, we have
\begin{align*}
\lim_{s\uparrow w(\beta)}g_0(s)
&=
-\frac{1-w(\beta)}{\beta-\alpha}
\lim_{s\uparrow w(\beta)}
(w^{-1}(s))'
<0,
\end{align*}
and
\begin{align*}
\lim_{s\downarrow w(\alpha)}g_0(s)
&=
1-
\frac{1-w(\alpha)}{\beta-\alpha}
\lim_{s\downarrow w(\alpha)}
(w^{-1}(s))'
>0.
\notag
\end{align*}
Therefore, there exists a unique
$t_0\in(w(\alpha),w(\beta))$ such that $g_0(t_0)=0$.
Using the same argument as in the proof of Lemma
\ref{lem3.7}(1), we can prove that $\delta_0(s)$ is the convex
envelope of $\varphi_0(s)$.

\item[\bf{(4)}]
Similarly, using the same argument as in the proof of Lemma
\ref{lem3.7}(1), once the contact points are shown to exist uniquely,
we can prove that $\delta_\lambda(s)$ is the convex envelope of
$\varphi_\lambda(s)$. We only need to prove the existence and
uniqueness of
\[
w(\alpha)<t_1(\lambda)<w(\beta)<t_2(\lambda)<1
\]
satisfying
\begin{align}\label{B.2}
\begin{cases}
\varphi_\lambda'(t_1(\lambda))
=
\varphi_\lambda'(t_2(\lambda)),
\\
\displaystyle
\varphi_\lambda'(t_1(\lambda))
=
\frac{
\varphi_\lambda(t_2(\lambda))
-
\varphi_\lambda(t_1(\lambda))
}{
t_2(\lambda)-t_1(\lambda)
}.
\end{cases}
\end{align}
Let
\begin{align}
\underline{s}(\lambda)
=
w\left(
1-F_\xi\left(
\frac{A_\lambda}{\lambda}
G_\xi(1-\beta)
\right)
\right).
\notag
\end{align}
Due to the strict monotonicity of $w$ and $F_\xi$, together with
$G_\xi>0$, we have
\begin{align}
\underline{s}(\lambda)
<
w\left(
1-F_\xi(G_\xi(1-\beta))
\right)
=
w(\beta).
\notag
\end{align}

By the original setup and monotonicity, for
$\underline{s}(\lambda)<s<w(\beta)$,
\begin{align}
y(s)
&=
w\left(
1-F_\xi\left(
\frac{\lambda}{A_\lambda}
G_\xi(1-w^{-1}(s))
\right)
\right)
\notag\\
&>
w\left(
1-F_\xi\left(
\frac{\lambda}{A_\lambda}
G_\xi(1-w^{-1}(\underline{s}(\lambda)))
\right)
\right)
\notag\\
&=
w\left(
1-F_\xi\left(
\frac{\lambda}{A_\lambda}
G_\xi\left(
F_\xi\left(
\frac{A_\lambda}{\lambda}
G_\xi(1-\beta)
\right)
\right)
\right)
\right)
\notag\\
&=
w\left(
1-F_\xi(G_\xi(1-\beta))
\right)
\notag\\
&=
w(\beta).
\notag
\end{align}
Moreover, from the definition of $y(s)$,
\begin{align}
\varphi_\lambda'(s)
=
\varphi_\lambda'(y(s)).
\label{B.3}
\end{align}

Let
\begin{align}
h_\lambda(s)
=
\varphi_\lambda(y(s))
-
\varphi_\lambda(s)
-
\varphi_\lambda'(s)(y(s)-s),
\qquad
w(\alpha)<s<w(\beta).
\notag
\end{align}
For
$\underline{s}(\lambda)<s_1<s_2<w(\beta)$, using integration by
parts, we obtain
\[
h_\lambda(s_1)-h_\lambda(s_2)>0,
\]
so $h_\lambda(s)$ is a strictly decreasing function of $s$.

Furthermore, due to the strict convexity of $\varphi_\lambda$ on
$(w(\alpha),w(\beta))$ and $(w(\beta),1)$, and
\[
\lim_{s\downarrow\underline{s}(\lambda)}y(s)=w(\beta),
\]
we have
\begin{align}
\lim_{s\downarrow\underline{s}(\lambda)}
h_\lambda(s)
&=
\int_{\underline{s}(\lambda)}^{w(\beta)}
\varphi_\lambda'(z)\dz
-
\varphi_\lambda'(\underline{s}(\lambda))
\bigl(w(\beta)-\underline{s}(\lambda)\bigr)
\notag\\
&=
\int_{\underline{s}(\lambda)}^{w(\beta)}
\left(
\varphi_\lambda'(z)
-
\varphi_\lambda'(\underline{s}(\lambda))
\right)\dz
>0,
\notag
\end{align}
and
\begin{align}
\lim_{s\uparrow w(\beta)}
h_\lambda(s)
=
\int_{w(\beta)}^{y(w(\beta)-)}
\left(
\varphi_\lambda'(z)
-
\varphi_\lambda'(y(w(\beta)-))
\right)\dz
<0.
\notag
\end{align}
Therefore, there exists a unique
$t_1(\lambda)\in(\underline{s}(\lambda),w(\beta))$ such that
\[
h_\lambda(t_1(\lambda))=0.
\]
Let
\[
t_2(\lambda)=y(t_1(\lambda)).
\]
Then
\[
w(\alpha)<t_1(\lambda)<w(\beta)<t_2(\lambda)<1.
\]
By \eqref{B.3},
\[
\varphi_\lambda'(t_1(\lambda))
=
\varphi_\lambda'(t_2(\lambda)),
\]
while $h_\lambda(t_1(\lambda))=0$ gives
\[
\varphi_\lambda'(t_1(\lambda))
=
\frac{
\varphi_\lambda(t_2(\lambda))
-
\varphi_\lambda(t_1(\lambda))
}{
t_2(\lambda)-t_1(\lambda)
}.
\]
Hence $t_1(\lambda)$ and $t_2(\lambda)$ satisfy
\eqref{B.2}. The strict monotonicity of $h_\lambda$ implies the uniqueness of
$t_1(\lambda)$. Since $t_2(\lambda)=y(t_1(\lambda))$ and $y$ is
single-valued under Assumption~\ref{ass:regularity-xi}, $t_2(\lambda)$
is unique as well.
\end{enumerate}

\section{Proofs of Propositions \ref{pro3.8} and \ref{pro3.10}}
\label{C.1}

Recall that
\begin{align}
\varphi_\lambda(s)=
\begin{cases}
\lambda w^{-1}(s), & s\in[0,w(\alpha)],\\
\lambda w^{-1}(s)+(1-\lambda)\dfrac{w^{-1}(s)-\alpha}{\beta-\alpha}, & s\in(w(\alpha),w(\beta)],\\
\lambda w^{-1}(s)+1-\lambda, & s\in(w(\beta),1].
\end{cases}
\label{C.1a}
\end{align}
Moreover,
\begin{align}
X_{T,\lambda}^{\mathrm{RVaR}}=\frac{x}{\mathbb E[\xi_T]}\delta_\lambda'\!\left(w(1-F_\xi(\xi_T))\right),
\label{C.2}
\end{align}
and
\begin{align}
(w^{-1}(s))'=\frac{\mathbb E[\xi_T]}{G_\xi(1-w^{-1}(s))}.
\label{C.3}
\end{align}

We first prove Proposition \ref{pro3.8}.

\begin{enumerate}

\item[\bf{1)}]
When $\lambda=0$, Lemma \ref{lem3.7}(1) gives
\begin{align}
\delta_0'(s)=
\begin{cases}
0, & s\in[0,w(\alpha)],\\
\dfrac{1}{1-w(\alpha)}, & s\in(w(\alpha),1].
\end{cases}
\notag
\end{align}
Since $w$ is strictly increasing,
\begin{align}
w(1-F_\xi(\xi_T))\le w(\alpha)\quad\Longleftrightarrow\quad \xi_T\ge\xi_{1-\alpha},\qquad \xi_{1-\alpha}=G_\xi(1-\alpha).
\notag
\end{align}
Substituting into \eqref{C.2} yields
\begin{align}
X_{T,0}^{\mathrm{RVaR}}=
\begin{cases}
0, & \xi_T>\xi_{1-\alpha},\medskip\\
X_{\mathrm{RVaR}}^{(0)}, & 0<\xi_T\le\xi_{1-\alpha},
\end{cases}
\qquad
X_{\mathrm{RVaR}}^{(0)}=\frac{x}{\mathbb E[\xi_T]\,[1-w(\alpha)]}.
\notag
\end{align}
This proves Proposition \ref{pro3.8}(1).

\item[\bf{2)}]
When $0<\lambda<1$, Lemma \ref{lem3.7}(2) gives
\begin{align}
\delta_\lambda'(s)=
\begin{cases}
\varphi_\lambda'(s), & s\in[0,w(\alpha)],\medskip\\
\varphi_\lambda'(t_0(\lambda)), & s\in(w(\alpha),t_0(\lambda)],\medskip\\
\varphi_\lambda'(s), & s\in(t_0(\lambda),1].
\end{cases}
\notag
\end{align}
On $[0,w(\alpha)]\cup(w(\beta),1]$, \eqref{C.1a} and \eqref{C.3} imply
\begin{align}
\varphi_\lambda'(s)=\frac{\lambda\mathbb E[\xi_T]}{G_\xi(1-w^{-1}(s))}.
\notag
\end{align}
Define
\begin{align}
\widehat\xi_{\mathrm{RVaR}}=G_\xi\!\left(1-w^{-1}(t_0(\lambda))\right).
\notag
\end{align}
Since $t_0(\lambda)>w(\beta)$,
\begin{align}
\frac{x}{\mathbb E[\xi_T]}\varphi_\lambda'(t_0(\lambda))=\frac{\lambda x}{\widehat\xi_{\mathrm{RVaR}}}=\widehat X_{\mathrm{RVaR}}.
\notag
\end{align}
Therefore,
\begin{align}
X_{T,\lambda}^{\mathrm{RVaR}}=
\begin{cases}
\dfrac{\lambda x}{\xi_T}, & \xi_T>\xi_{1-\alpha},\medskip\\
\widehat X_{\mathrm{RVaR}}, & \widehat\xi_{\mathrm{RVaR}}<\xi_T\le\xi_{1-\alpha},\medskip\\
\dfrac{\lambda x}{\xi_T}, & 0<\xi_T\le\widehat\xi_{\mathrm{RVaR}}.
\end{cases}
\notag
\end{align}
This proves Proposition \ref{pro3.8}(2).

\end{enumerate}

We next prove Proposition \ref{pro3.10}.

\begin{enumerate}

\item[\bf{1)}]
When $\lambda=0$, Lemma \ref{lem3.7}(3) gives
\begin{align}
\delta_0'(s)=
\begin{cases}
\varphi_0'(s), & s\in[0,t_0],\medskip\\
\varphi_0'(t_0), & s\in(t_0,1].
\end{cases}
\notag
\end{align}
From \eqref{C.1a} and \eqref{C.3},
\begin{align}
\varphi_0'(s)=\frac{\mathbb{E}[\xi_T]}{(\beta-\alpha)G_\xi(1-w^{-1}(s))},\qquad s\in(w(\alpha),w(\beta)).
\notag
\end{align}
Define
\begin{align}
\widetilde\xi_{\mathrm{RVaR}}=G_\xi\!\left(1-w^{-1}(t_0)\right),\qquad \widetilde X_{\mathrm{RVaR}}=\frac{x}{(\beta-\alpha)\widetilde\xi_{\mathrm{RVaR}}}.
\notag
\end{align}
Substituting into \eqref{C.2} yields
\begin{align}
X_{T,0}^{\mathrm{RVaR}}=
\begin{cases}
0, & \xi_T>\xi_{1-\alpha},\medskip\\
\dfrac{x}{(\beta-\alpha)\xi_T}, & \widetilde\xi_{\mathrm{RVaR}}<\xi_T\le\xi_{1-\alpha},\medskip\\
\widetilde X_{\mathrm{RVaR}}, & 0<\xi_T\le\widetilde\xi_{\mathrm{RVaR}}.
\end{cases}
\notag
\end{align}
This proves Proposition \ref{pro3.10}(1).

\item[\bf{2)}]
When $0<\lambda<1$, Lemma \ref{lem3.7}(4) gives
\begin{align}
\delta_\lambda'(s)=
\begin{cases}
\varphi_\lambda'(s), & s\in[0,t_1(\lambda)],\medskip\\
\varphi_\lambda'(t_1(\lambda)), & s\in(t_1(\lambda),t_2(\lambda)],\medskip\\
\varphi_\lambda'(s), & s\in(t_2(\lambda),1].
\end{cases}
\notag
\end{align}
Using \eqref{C.1a} and \eqref{C.3},
\begin{align}
\varphi_\lambda'(s)=
\begin{cases}
\dfrac{\lambda\mathbb E[\xi_T]}{G_\xi(1-w^{-1}(s))}, & s\in[0,w(\alpha)]\cup(w(\beta),1],\medskip\\
\dfrac{A_\lambda\mathbb E[\xi_T]}{G_\xi(1-w^{-1}(s))}, & s\in(w(\alpha),w(\beta)].
\end{cases}
\notag
\end{align}
Define
\begin{align}
\bar\xi_{\mathrm{RVaR}}=G_\xi\!\left(1-w^{-1}(t_1(\lambda))\right).
\notag
\end{align}
Since $\varphi_\lambda'(t_1(\lambda))=\varphi_\lambda'(t_2(\lambda))$,
\begin{align}
A_\lambda\frac{\mathbb E[\xi_T]}{\bar\xi_{\mathrm{RVaR}}}=\frac{\lambda\mathbb E[\xi_T]}{G_\xi(1-w^{-1}(t_2(\lambda)))}.
\notag
\end{align}
Hence
\begin{align}
\underline\xi_{\mathrm{RVaR}}=G_\xi\!\left(1-w^{-1}(t_2(\lambda))\right)=\frac{\lambda}{A_\lambda}\bar\xi_{\mathrm{RVaR}}=\frac{\lambda(\beta-\alpha)}{1-\lambda+\lambda(\beta-\alpha)}\bar\xi_{\mathrm{RVaR}}.
\notag
\end{align}
Moreover,
\begin{align}
\underline X_{\mathrm{RVaR}}=\frac{x}{\mathbb E[\xi_T]}\varphi_\lambda'(t_1(\lambda))=\frac{A_\lambda x}{\bar\xi_{\mathrm{RVaR}}}=\frac{\lambda x}{\underline\xi_{\mathrm{RVaR}}}.
\notag
\end{align}
Therefore,
\begin{align}
X_{T,\lambda}^{\mathrm{RVaR}}=
\begin{cases}
\dfrac{\lambda x}{\xi_T}, & \xi_T\ge\xi_{1-\alpha},\medskip\\
A_\lambda\dfrac{x}{\xi_T}, & \bar\xi_{\mathrm{RVaR}}\le\xi_T<\xi_{1-\alpha},\medskip\\
\underline X_{\mathrm{RVaR}}, & \underline\xi_{\mathrm{RVaR}}\le\xi_T<\bar\xi_{\mathrm{RVaR}},\medskip\\
\dfrac{\lambda x}{\xi_T}, & 0<\xi_T<\underline\xi_{\mathrm{RVaR}}.
\end{cases}
\notag
\end{align}
This proves Proposition \ref{pro3.10}(2).

\end{enumerate}

\section{Proofs of Corollaries \ref{cor3.9} and \ref{cor3.11}}
\label{D.1}

Conditional on $\mathcal F_t$, $\ln\xi_T$ is normally distributed with mean
$\ln\xi_t-(r+\frac{1}{2}|\theta|^2)(T-t)$ and variance $|\theta|^2(T-t)$.
Hence, for any $y>0$,
\begin{align}
\mathbb P(\xi_T\le y\mid\mathcal F_t)
&=N(d_1(t,\xi_t,y)), \qquad
\mathbb P(\xi_T>y\mid\mathcal F_t)
=N(-d_1(t,\xi_t,y)).
\label{D.1a}
\end{align}
Moreover,
\begin{align}
\frac{1}{\xi_t}\mathbb E[\xi_T\mathbf 1_{\{\xi_T\le y\}}\mid\mathcal F_t]
=e^{-r(T-t)}N(d_2(t,\xi_t,y)).
\label{D.2}
\end{align}
For any attainable terminal wealth $X_T$, the corresponding wealth process satisfies
\begin{align}
X_t=\frac{1}{\xi_t}\mathbb E[\xi_TX_T\mid\mathcal F_t].
\label{D.3}
\end{align}
If $X_t=f(t,\xi_t)$, then It\^o's formula and
$\dd\xi_t=-r\xi_t\dt-\xi_t\theta^\top \dw_t$ imply
\begin{align}
\pi_t=-\xi_t f_\xi(t,\xi_t)(\sigma_t^\top)^{-1}\theta.
\label{D.4}
\end{align}
We also use the identity
\begin{align}
\nu(d_1(t,\xi_t,y))
=e^{-r(T-t)}\frac{\xi_t}{y}\nu(d_2(t,\xi_t,y)).
\label{D.5}
\end{align}

We first prove Corollary \ref{cor3.9}.

\begin{enumerate}

\item[\bf{1)}]
When $\lambda=0$, Proposition \ref{pro3.8}(1) gives
\begin{align}
X_{T,0}^{\mathrm{RVaR}}
=X_{\mathrm{RVaR}}^{(0)}
\mathbf 1_{\{\xi_T\le\xi_{1-\alpha}\}}.
\notag
\end{align}
By \eqref{D.2}--\eqref{D.3},
\begin{align}
X_{t,0}^{\mathrm{RVaR}}
=X_{\mathrm{RVaR}}^{(0)}e^{-r(T-t)}
N(d_2(t,\xi_t,\xi_{1-\alpha})).
\notag
\end{align}
Applying \eqref{D.4} yields
\begin{align}
\pi_{t,0}^{\mathrm{RVaR}}
=
\frac{
X_{\mathrm{RVaR}}^{(0)}e^{-r(T-t)}
\nu(d_2(t,\xi_t,\xi_{1-\alpha}))
}{
|\theta|\sqrt{T-t}
}
(\sigma_t^\top)^{-1}\theta.
\notag
\end{align}

\item[\bf{2)}]
When $0<\lambda<1$, Proposition \ref{pro3.8}(2) gives
\begin{align}
X_{T,\lambda}^{\mathrm{RVaR}}
={}&
\frac{\lambda x}{\xi_T}
\mathbf 1_{\{\xi_T>\xi_{1-\alpha}\}}
+
\widehat X_{\mathrm{RVaR}}
\mathbf 1_{\{\widehat\xi_{\mathrm{RVaR}}<\xi_T\le\xi_{1-\alpha}\}} +
\frac{\lambda x}{\xi_T}
\mathbf 1_{\{\xi_T\le\widehat\xi_{\mathrm{RVaR}}\}}.
\notag
\end{align}
Using \eqref{D.1a}--\eqref{D.3}, we obtain
\begin{align}
X_{t,\lambda}^{\mathrm{RVaR}}
={}&
\frac{\lambda x}{\xi_t}
\left[
N(-d_1(t,\xi_t,\xi_{1-\alpha}))
+
N(d_1(t,\xi_t,\widehat\xi_{\mathrm{RVaR}}))
\right]
\notag\\
&+
\widehat X_{\mathrm{RVaR}}e^{-r(T-t)}
\left[
N(d_2(t,\xi_t,\xi_{1-\alpha}))
-
N(d_2(t,\xi_t,\widehat\xi_{\mathrm{RVaR}}))
\right].
\notag
\end{align}
Since $\widehat X_{\mathrm{RVaR}}=\lambda x/\widehat\xi_{\mathrm{RVaR}}$,
\eqref{D.4}--\eqref{D.5} give
\begin{align}
\pi_{t,\lambda}^{\mathrm{RVaR}}
={}&
\Bigg\{
\frac{\lambda x}{\xi_t}
\left[
N(-d_1(t,\xi_t,\xi_{1-\alpha}))
+
N(d_1(t,\xi_t,\widehat\xi_{\mathrm{RVaR}}))
\right]
\notag\\
&\quad+
\frac{
e^{-r(T-t)}
\nu(d_2(t,\xi_t,\xi_{1-\alpha}))
}{
|\theta|\sqrt{T-t}
}
\left(
\widehat X_{\mathrm{RVaR}}
-
\frac{\lambda x}{\xi_{1-\alpha}}
\right)
\Bigg\}
(\sigma_t^\top)^{-1}\theta.
\notag
\end{align}

\end{enumerate}

We next prove Corollary \ref{cor3.11}.

\begin{enumerate}

\item[\bf{1)}]
When $\lambda=0$, Proposition \ref{pro3.10}(1) gives
\begin{align}
X_{T,0}^{\mathrm{RVaR}}
={}&
\frac{x}{(\beta-\alpha)\xi_T}
\mathbf 1_{\{\widetilde\xi_{\mathrm{RVaR}}<\xi_T\le\xi_{1-\alpha}\}} +
\widetilde X_{\mathrm{RVaR}}
\mathbf 1_{\{\xi_T\le\widetilde\xi_{\mathrm{RVaR}}\}}.
\notag
\end{align}
By \eqref{D.1a}--\eqref{D.3},
\begin{align}
X_{t,0}^{\mathrm{RVaR}}
={}&
\frac{x}{(\beta-\alpha)\xi_t}
\left[
N(d_1(t,\xi_t,\xi_{1-\alpha}))
-
N(d_1(t,\xi_t,\widetilde\xi_{\mathrm{RVaR}}))
\right]
\notag\\
&+
\widetilde X_{\mathrm{RVaR}}e^{-r(T-t)}
N(d_2(t,\xi_t,\widetilde\xi_{\mathrm{RVaR}})).
\notag
\end{align}
Since
$\widetilde X_{\mathrm{RVaR}}
=x/[(\beta-\alpha)\widetilde\xi_{\mathrm{RVaR}}]$,
\eqref{D.4}--\eqref{D.5} yield
\begin{align}
\pi_{t,0}^{\mathrm{RVaR}}
={}&
\frac{x}{(\beta-\alpha)\xi_t}
\Bigg[
N(d_1(t,\xi_t,\xi_{1-\alpha}))
-
N(d_1(t,\xi_t,\widetilde\xi_{\mathrm{RVaR}}))
\notag\\
&\qquad\qquad\quad+
\frac{
\nu(d_1(t,\xi_t,\xi_{1-\alpha}))
}{
|\theta|\sqrt{T-t}
}
\Bigg]
(\sigma_t^\top)^{-1}\theta.
\notag
\end{align}

\item[\bf{2)}]
When $0<\lambda<1$, Proposition \ref{pro3.10}(2) gives
\begin{align}
X_{T,\lambda}^{\mathrm{RVaR}}
={}&
\frac{\lambda x}{\xi_T}
\mathbf 1_{\{\xi_T\ge\xi_{1-\alpha}\}}
+
\frac{A_\lambda x}{\xi_T}
\mathbf 1_{\{\bar\xi_{\mathrm{RVaR}}\le\xi_T<\xi_{1-\alpha}\}}
\notag\\
&+
\underline X_{\mathrm{RVaR}}
\mathbf 1_{\{\underline\xi_{\mathrm{RVaR}}\le\xi_T<\bar\xi_{\mathrm{RVaR}}\}}
+
\frac{\lambda x}{\xi_T}
\mathbf 1_{\{\xi_T<\underline\xi_{\mathrm{RVaR}}\}}.
\notag
\end{align}
Using \eqref{D.1a}--\eqref{D.3}, we obtain
\begin{align}
X_{t,\lambda}^{\mathrm{RVaR}}
={}&
\frac{\lambda x}{\xi_t}
N(-d_1(t,\xi_t,\xi_{1-\alpha}))
\notag\\
&+
\frac{A_\lambda x}{\xi_t}
\left[
N(d_1(t,\xi_t,\xi_{1-\alpha}))
-
N(d_1(t,\xi_t,\bar\xi_{\mathrm{RVaR}}))
\right]
\notag\\
&+
\underline X_{\mathrm{RVaR}}e^{-r(T-t)}
\left[
N(d_2(t,\xi_t,\bar\xi_{\mathrm{RVaR}}))
-
N(d_2(t,\xi_t,\underline\xi_{\mathrm{RVaR}}))
\right]
\notag\\
&+
\frac{\lambda x}{\xi_t}
N(d_1(t,\xi_t,\underline\xi_{\mathrm{RVaR}})).
\notag
\end{align}
Using
\[
\underline X_{\mathrm{RVaR}}
=
A_\lambda\frac{x}{\bar\xi_{\mathrm{RVaR}}}
=
\lambda\frac{x}{\underline\xi_{\mathrm{RVaR}}},
\]
the boundary terms at $\bar\xi_{\mathrm{RVaR}}$ and
$\underline\xi_{\mathrm{RVaR}}$ cancel. Hence,
\begin{align}
\pi_{t,\lambda}^{\mathrm{RVaR}}
={}&
\frac{x}{\xi_t}
\Bigg\{
\lambda N(-d_1(t,\xi_t,\xi_{1-\alpha}))
\notag\\
&\qquad+
A_\lambda
\left[
N(d_1(t,\xi_t,\xi_{1-\alpha}))
-
N(d_1(t,\xi_t,\bar\xi_{\mathrm{RVaR}}))
\right]
\notag\\
&\qquad+
\lambda N(d_1(t,\xi_t,\underline\xi_{\mathrm{RVaR}}))
+
(A_\lambda-\lambda)
\frac{
\nu(d_1(t,\xi_t,\xi_{1-\alpha}))
}{
|\theta|\sqrt{T-t}
}
\Bigg\}
(\sigma_t^\top)^{-1}\theta.
\notag
\end{align}

\end{enumerate}
 


\begin{thebibliography}{}

\bibitem[Acerbi, 2002]{Acerbi}
Acerbi, C. (2002).
\newblock {Spectral measures of risk: A coherent representation of subjective
  risk aversion}.
\newblock {\em Journal of Banking \& Finance}, 26(7), 1505--1518.

\bibitem[Artzner et~al., 1999]{Artzner}
Artzner, P., Delbaen, F., Eber, J., \& Heath, D. (1999).
\newblock {Coherent measures of risk}.
\newblock {\em Mathematical Finance}, 9, 203--228.

\bibitem[Basak \& Chabakauri, 2010]{Basak}
Basak, S. \& Chabakauri, G. (2010).
\newblock {Dynamic mean-variance asset allocation}.
\newblock {\em The Review of Financial Studies}, 23(8), 2970--3016.

\bibitem[Cont et~al., 2010]{Cont}
Cont, R., Deguest, R., \& Scandolo, G. (2010).
\newblock {Robustness and sensitivity analysis of risk measurement procedures}.
\newblock {\em Quantitative Finance}, 10(6), 593--606.

\bibitem[Danielsson et~al., 2005]{Danielsson}
Danielsson, J., Jorgensen, B., Samorodnitsky, G., Sarma, M., \& de~Vries, C.
  (2005).
\newblock {\em {Subadditivity re-examined: The case for Value-at-Risk,
  Preprint}}.
\newblock London School of Economics.

\bibitem[Embrechts et~al., 2018]{Embrechts}
Embrechts, P., Liu, H., \& Wang, R. (2018).
\newblock {Quantile-based risk sharing}.
\newblock {\em Operations Research}, 66(4), 936--949.

\bibitem[He \& Jiang, 2022]{He1}
He, X.~D. \& Jiang, Z. (2022).
\newblock {Mean-variance portfolio selection with dynamic targets for expected
  terminal wealth}.
\newblock {\em Mathematics of Operations Research}, 47(1), 587--615.

\bibitem[He et~al., 2015]{He2}
He, X.~D., Jin, H., \& Zhou, X.~Y. (2015).
\newblock {Dynamic portfolio choice when risk is measured by weighted VaR}.
\newblock {\em Mathematics of Operations Research}, 40(3), 773--796.

\bibitem[He \& Zhou, 2011]{He3}
He, X.~D. \& Zhou, X.~Y. (2011).
\newblock {Portfolio choice via quantiles}.
\newblock {\em Mathematical Finance}, 21(2), 203--231.

\bibitem[Ibragimov \& Walden, 2007]{Ibragimov}
Ibragimov, R. \& Walden, J. (2007).
\newblock {The limits of diversification when losses may be large}.
\newblock {\em Journal of Banking and Finance}, 31, 2551--2569.

\bibitem[Jin et~al., 2006]{Jin1}
Jin, H., Markowitz, H., \& Zhou, X.~Y. (2006).
\newblock {A note on semivariance}.
\newblock {\em Mathematical Finance}, 16(1), 53--61.

\bibitem[Jin et~al., 2005]{Jin2}
Jin, H., Yan, J.-A., \& Zhou, X.~Y. (2005).
\newblock {Continuous-time mean--risk portfolio selection}.
\newblock {\em Annales de l'Institut Henri Poincar\'e (B) Probability and
  Statistics}, volume~41, 559--580.

\bibitem[Jin \& Zhou, 2008]{Jin3}
Jin, H. \& Zhou, X.~Y. (2008).
\newblock {Behavioral portfolio selection in continuous time}.
\newblock {\em Mathematical Finance}, 18(3), 385--426.

\bibitem[Karatzas \& Shreve, 1998]{Karatzas}
Karatzas, I. \& Shreve, S.~E. (1998).
\newblock {\em {Methods of Mathematical Finance}}, volume~39.
\newblock Springer Science \& Business Media.

\bibitem[Kelly, 1956]{Kelly}
Kelly, J. (1956).
\newblock {A new interpretation of information rate}.
\newblock {\em Bell System Technical Journal}, 35, 917--926.

\bibitem[Kou \& Peng, 2016]{Kou}
Kou, S. \& Peng, X. (2016).
\newblock {On the measurement of economic tail risk}.
\newblock {\em Operations Research}, 64, 1056--1072.

\bibitem[Li et~al., 2018]{Li}
Li, L.~J., Shao, H., Wang, R.~D., \& Yang, J.~P. (2018).
\newblock {Worst-case range value-at-risk with partial information}.
\newblock {\em SIAM Journal on Financial Mathematics}, 9(1), 190--218.

\bibitem[Maria~M\"uller et~al., 2024]{Maria}
Maria~M\"uller, F., Weber~G\"ossling, T., Solgon~Santos, S., \& Brutti~Righi,
  M. (2024).
\newblock {A comparison of Range Value at Risk (RVaR) forecasting models}.
\newblock {\em Journal of Forecasting}, 43(3), 509--543.

\bibitem[Markowitz, 1952]{Markowitz}
Markowitz, H. (1952).
\newblock {Portfolio selection}.
\newblock {\em The Journal of Finance}, 7(1), 77--91.

\bibitem[Rockafellar \& Uryasev, 2000]{Rockafellar1}
Rockafellar, R.~T. \& Uryasev, S. (2000).
\newblock {Optimization of conditional value-at-risk}.
\newblock {\em Journal of Risk}, 2, 21--42.

\bibitem[Rockafellar \& Uryasev, 2002]{Rockafellar2}
Rockafellar, R.~T. \& Uryasev, S. (2002).
\newblock {Conditional value-at-risk for general loss distributions}.
\newblock {\em Journal of Banking \& Finance}, 26(7), 1443--1471.

\bibitem[Wang \& Wei, 2020]{Wang}
Wang, R.~D. \& Wei, Y.~R. (2020).
\newblock {Characterizing optimal allocations in quantile-based risk sharing}.
\newblock {\em Insurance: Mathematics and Economics}, 93, 288--300.

\bibitem[Wei, 2018]{Wei1}
Wei, P. (2018).
\newblock {Risk management with weighted VaR}.
\newblock {\em Mathematical Finance}, 28(4), 1020--1060.

\bibitem[Wei, 2021]{Wei2}
Wei, P. (2021).
\newblock {Risk management with expected shortfall}.
\newblock {\em Mathematics and Financial Economics}, 15(4), 847--883.

\bibitem[Wei \& Xu, 2025]{Wei3}
Wei, P. \& Xu, Z.~Q. (2025).
\newblock {Dynamic growth-optimal portfolio choice under risk control}.
\newblock {\em European Journal of Operational Research}, 322(1), 325--340.

\bibitem[Xu, 2016]{Xu2}
Xu, Z.~Q. (2016).
\newblock {A note on the quantile formulation}.
\newblock {\em Mathematical Finance}, 26(3), 589--601.

\bibitem[Zhang \& Meng, 2022]{Zhang}
Zhang, W. \& Meng, H. (2022).
\newblock {Robust optimal dynamic reinsurance policies under the mean-RVaR
  premium principle}.
\newblock {\em Communications in Statistics - Theory and Methods}, 53(1),
  113--143.

\end{thebibliography}
\end{document}